\documentclass[twocolumn,preprintnumbers,amsmath,amssymb,prc]{revtex4}

\usepackage{graphicx}

\usepackage{dcolumn}
\usepackage{color}
\usepackage{bm}

\usepackage{algorithm}
\usepackage{algpseudocode}

\begin{document}

\title{
  Thick-Restart Block Lanczos Method for
  Large-Scale Shell-Model Calculations
  % Efficiency of Thick-Restart Block Lanczos Method in 
  % Large-Scale Shell-Model Calculations
}

\author{
  Noritaka Shimizu$^{1,}$\footnote{shimizu@cns.s.u-tokyo.ac.jp},
  Takahiro Mizusaki$^2$,
  Yutaka Utsuno$^{3,1}$  
  and Yusuke Tsunoda$^1$}

\affiliation{
  $^1$Center for Nuclear Study, The University of Tokyo, 
  7-3-1 Hongo, Bunkyo-ku, Tokyo 113-0033, Japan\\
  $^2$Institute of Natural Sciences, Senshu University, 
  3-8-1 Kanda-Jinbocho, Chiyoda-ku, Tokyo 101-8425, Japan\\
  $^3$Advanced Science Research Center, 
  Japan Atomic Energy Agency, Tokai, Ibaraki 319-1195, Japan 
}

\date{\today}

\begin{abstract}
  We propose a thick-restart block Lanczos method, 
  which is an extension of the thick-restart Lanczos method
  with the block algorithm, as an eigensolver 
  of the large-scale shell-model calculations.
  This method has two advantages over the conventional Lanczos method:
  the precise computations of the near-degenerate eigenvalues, 
  and the efficient computations 
  for obtaining a large number of eigenvalues.
  These features are quite advantageous 
  to compute highly excited states where
  the eigenvalue density is rather high.
  A shell-model code, named KSHELL, equipped with this method 
  was developed for massively parallel computations, 
  and it enables us to reveal nuclear statistical properties
  which are intensively investigated by recent experimental
  facilities.
  We describe the algorithm and performance of the KSHELL code
  and demonstrate that the present method outperforms
  the conventional Lanczos method.
\end{abstract}

\maketitle

\section{Introduction}
\label{sec:intro}

Solving a quantum many-body problem 
having protons and neutrons as constituent particles
is one of the ultimate goals in nuclear structure physics.
Although nucleons do not have an external field 
like electrons in an atom, nuclear shell model is 
successful in describing the low-lying excitation 
spectra of nuclei near closed-shell nuclei \cite{mayerjensen}.
Based on the success of the nuclear shell model, 
large-scale shell-model (LSSM) calculations
have been performed to go far 
beyond closed-shell nuclei.
In the LSSM, 
we assume that a nucleus is composed 
of an inert core and active particles 
that move in some active orbitals. 
The active particles and active orbitals 
are usually taken as valence particles and the orbitals in the valence shell,
respectively.
The nuclear wave function is expressed as
a superposition of the Slater determinants,
which represent occupations of the active particles
in the orbitals. 
The LSSM is also called configuration interaction calculations 
like in quantum chemistry. 
By utilizing the LSSM, low-energy nuclear spectroscopic data
of $sd$-shell \cite{usd, usdab} and
$pf$-shell \cite{gxpf1a,jun45} nuclei have been investigated
systematically.
Recent progress in radioactive ion beam facilities
enables us to reveal exotic nuclear structures 
of unstable nuclei \cite{otsuka-review,caurier_rmp}.

In shell-model calculations we 
solve the Schr\"{o}dinger equation of protons and neutrons
as an eigenvalue problem 
of a huge sparse real symmetric matrix
utilizing the traditional methods: 
the Lanczos method \cite{lanczos} and the thick-restart
Lanczos method \cite{tr-lanczos}.
These methods are known to be quite effective
to obtain a small number of the lowest eigenvalues
of a sparse matrix.
Moreover, several efforts have been paid to
pursue a better eigensolver, 
such as the Sakurai-Sugiura method \cite{mizusaki-ss, ss-method} and the
Locally Optimal Block Preconditioned Conjugate Gradient
method \cite{mfdn-lobpcg}.

In many cases, a small number of low-lying eigenvalues
and eigenvectors 
need to be calculated by the LSSM calculations,
since in many experimental studies of unstable nuclei
only a small number of low-lying states can be measured.
In order to analyze such low-lying states, more
than a dozen shell-model codes had been developed
\cite{antoine,bigstick,eicode,kshell,nathan,nushell,nushellx,
MFDn,mshell,mshell64,oxbash, vecsse}.
However, recent progress of the experimental techniques 
extends the opportunity to investigate highly excited states 
and their statistic properties, such as $\gamma$-ray strength functions
and level densities \cite{oslo}.
In order to discuss these properties by shell-model calculations precisely, 
a relatively large number of eigenstates ($O(10^2)$-$O(10^3)$)
are required by solving the eigenvalue problem
\cite{La133-palit,sieja-leh,jorgen-m1,ld-ni58}.
In the present paper, 
we propose the thick-restart block
Lanczos method 
to compute these states efficiently,
and describe the implementation and the performance of the
KSHELL code which we developed \cite{kshell}.
In the code we adopt an algorithm
of generating the matrix elements on the fly
in order to avoid storing the matrix elements and
to save memory usage, while 
the generation of the matrix elements
costs a certain amount of the computation time.
We will also demonstrate the block method reduces this cost
in the present paper.

The thick-restart block Lanczos method
is a combination of the block Lanczos method 
and the thick-restart method.
The block Lanczos method was proposed as 
a general eigenvalue solver by several authors
(e.g., \cite{block-lanc,cullem-block-lanc}).
% It has two advantages over the simple Lanczos method:
% One is that it correctly provides us
% with degenerate eigenvalues. 
% The other is that a product of a matrix
% and a single vector 
% is replaced by that of a matrix and a block of vectors,
% which causes efficient computation.
In comparison with the simple Lanczos method,
it is advantageous in that it enables us
to solve multiple eigenvalue problems and
it may be efficient for computing clustered eigenvalues
\cite{jia-block}.
% In addition, this method can 
% improve its computational perforamance in FLOPS
% for recent computers with slow memory access
% relative to the high-speed numerations. 
% In the method, a product of a matrix and a matrix, 
% which is bundled vectors, can be
% efficiently calculated than the corresponding
% products of a matrix and a vector.
In the present work, 
the block method is
expected to work more efficiently 
since the matrix elements are generated on the fly
in the KSHELL code and 
the cost of this generation can be
reduced by the block method.
In the block method, since the products of a matrix and multiple vectors
are performed at once, the frequency of the generations
of the matrix elements is reduced.
Moreover, the block method accelerates the convergence
of the Lanczos iterations
when a large number of the low-lying eigenvalues are required.

This paper is organized as follows.
The KSHELL code is based on the $M$-scheme representation
which is advantageous for large-scale calculations
and is discussed in Sect.~\ref{sec:lssm}.
The Lanczos method and its variants are discussed comparing with each other 
in Sect.~\ref{sec:lancs}.
Their performance in practical calculations are shown in 
Sect.~\ref{sec:perf}.
Sect.~\ref{sec:summary} concludes the paper.
We further describe the implementation of the KSHELL code
in the appendices. 
In Appx.~\ref{sec:part}, the $M$-scheme basis states
and its structure to be stored are discussed. 
The most time-consuming part of the algorithm
is the matrix-vector product appearing in the Lanczos algorithm.
The on-the-fly algorithm of the matrix-vector product 
is briefly described in Appx.~\ref{sec:matvec}.
In Appx.~\ref{sec:perf-parallel}, we discuss 
the way of the parallel computation of the matrix-vector product
and reorthogonalization, which are the most time-consuming parts of
the algorithm.

\section{LSSM with $M$-scheme basis states}
% \section{Framework of the LSSM calculations}
\label{sec:lssm}

In nuclear shell model calculations,
the shell-model wave function is
described as a superposition of configurations, which
represent various ways of the
occupation of active particles in the valence orbits.
Namely, the wave function is a linear combination of
a vast number of Slater determinants,
which are the antisymmetrized
products of the single-particle wave functions.
The simplest representation for a many-body Slater determinant
is called ``$M$-scheme'' basis state and described as 
\begin{equation}
    |M_i\rangle = 
    c^\dagger_{a_{i,1}}c^\dagger_{a_{i,2}} \cdots
    c^\dagger_{a_{i,A}}|-\rangle
  \label{eq:mbasis}
\end{equation}
where $A$ and $|-\rangle$ are the 
number of active nucleons and an inert core, respectively.
The $c^\dagger_{a_{i,1}}$ denotes a creation operator 
of the single-particle state $a_{i,1}$.
The Slater determinant $|M_i\rangle$
represents that the 1st, 2nd, $\cdots$, and $A$-th particles occupy
the $a_{i,1}, a_{i,2}, \cdots, $ and $a_{i,A}$
single-particle states, respectively.
The set $(a_{i,1}, a_{i,2}, a_{i,3}, \cdots , a_{i,A}) $
is sometimes called ``configuration''.
On computer programs, 
it is convenient to represent the configuration by
the bit representation 
with occupied and unoccupied states
being the bit 1 and bit 0.

Since the model space is fully spanned by 
the $M$-scheme basis states,
the shell-model wave function is expressed
as their linear combination,  
\begin{equation}
    \label{eq:wf}
    |\Psi \rangle = \sum_{i=1}^{D_M} v_i |M_i\rangle, 
\end{equation}
where the number of the $M$-scheme basis, $D_M$, is called 
the $M$-scheme dimension.
The coefficients $v_i$ are obtained by solving 
the Schr\"{o}dinger equation, or the eigenvalue problem,
in $M$-scheme basis as 
\begin{equation}
    \label{eq:schro-mat}
    \sum_{j=1}^{D_M} H_{ij} v_j = E v_i , 
\end{equation}
where $H_{ij} = \langle M_i|H|M_j\rangle $ is called 
the Hamiltonian matrix and is real symmetric.
Thus, the eigenvector $v_i$ contains all information 
of the shell-model wave function.

In usual shell-model calculations, the Hamiltonian 
consists of the one-body and two-body interactions as
\begin{equation}
  H = H^{(1)} + H^{(2)}= \sum_{ac} h^{(1)}_{ac} c^\dagger_a c_c
  + \sum_{a<b, c<d} h^{(2)}_{abcd} c^\dagger_a c^\dagger_b c_d c_c ,
\end{equation}
where $c^\dagger_a$ denotes a creation operator of
single-particle state $a$.
$H^{(1)}$ is a one-body Hamiltonian with the coefficients 
$h^{(1)}_{ac}$, 
whose diagonal part is the single-particle energy
of the single-particle orbit $a$.
$H^{(2)}$ is a two-body interaction which has
rotational and parity symmetries.
It is represented by the so-called Two-Body
Matrix Elements (TBMEs) \cite{ppnp_brown}.
In the present work, we do not discuss three-body interaction
which is often used in the no-core shell-model approach 
\cite{ncsm}.

The many-body Hamiltonian matrix $H_{ij}$ has
a block-diagonal structure thanks to the symmetries
of the Hamiltonian.
There are two symmetries which can be utilized in the $M$-scheme basis:
rotational symmetry around $z$-axis and parity symmetry.
The operators of the rotation and parity inversion
are referred by $J_z$ and $\Pi$
and the corresponding eigenvalues $M$ and $\pi$, respectively.
We only need to treat a block matrix specified by the $M$ and $\pi$.
In the case of even-mass nuclei, we need to construct a subspace
spanned by the Slater determinants 
only having $M=0$.
This subspace contains any $J$ states without duplication.
In the case of odd-mass nuclei, $M=\frac12$ subspace is
enough to obtain all the shell-model states. 
The $M$-scheme dimension usually denotes 
the largest dimension of such a block matrix.

\begin{figure}[htbp]
  \includegraphics[scale=0.4]{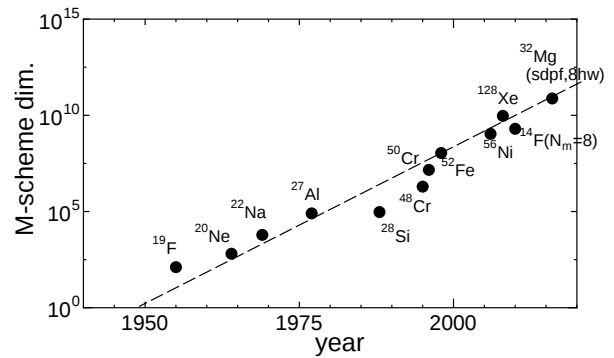}
  \caption{$M$-scheme dimension of the LSSM for nuclei
    as a function of the publication year.
    The line is drawn to guide the eyes.
    $^{14}$F ($N_m=8$) is taken from Ref.~\cite{14F}.
  }
  \label{fig:dim_year}
\end{figure}

Figure \ref{fig:dim_year} shows the historical progress of
the feasibility of the LSSM.
As computer performance grows, 
the tractable $M$-scheme dimension increases exponentially
as a function of the publication year and
recent maximum dimension reaches $10^{11}$.
The rightmost point in the figure, ``$^{32}$Mg (sdpf, 8hw)'', 
which means the LSSM of $^{32}$Mg with the $sd$- and $pf$-shells
model space allowing particle-hole excitations
up to $8\hbar\omega$ configurations,
was achieved by the KSHELL code \cite{ntsunoda-8hw}.

\section{Lanczos algorithms}
\label{sec:lancs}

Historically, the Lanczos method has been adopted
in many shell-model codes 
since pioneering works in 1970's \cite{M-lanczos,lanczos-sebe}
and continued to be utilized till now.
In practical calculations, a variation of the Lanczos methods, 
the thick-restart Lanczos method has often been 
adopted 
in order to reduce the elapsed time of 
the reorthogonalization  \cite{tr-lanczos}. 
Since the block Lanczos is used for efficient computations
to handle many eigenstates, 
we propose the thick-restart block Lanczos method for the LSSM
in this paper.
We briefly review the 
naive Lanczos method in Sect.~\ref{sec:lanc}, 
and thick-restart Lanczos method in Sect.~\ref{sec:trlan}.
The algorithms of 
the block Lanczos method and the thick-restart block Lanczos method 
are described in Sects.\ref{sec:blan} and \ref{sec:trblan}, respectively.

\subsection{Lanczos method}
\label{sec:lanc}

We begin with the simplest,
well-known Lanczos method \cite{lanczos}.
The Lanczos method is 
one of the most powerful methods
to obtain the lowest eigenstates of a sparse matrix.
In the algorithm, the eigenvalues of the Hamiltonian matrix, $H$, 
are approximated by the eigenvalues of the Krylov subspace \cite{krylov}
\begin{equation}
  { \cal  K}_{l_m}(H, \bm{v}_1) =
  \{\bm{v}_1, H\bm{v}_1, H^2\bm{v}_1, H^3\bm{v}_1,
  \cdots, H^{l_m-1}\bm{v}_{1} \}
\end{equation}
with $\bm{v}_1$ being an arbitrary initial vector.
The few lowest eigenvalues converge quite fast 
and the necessary number of iterations for convergence,
$l_m$, is much smaller than the dimension $D_M$ in general.
Moreover, since the $H$ appears only as a matrix-vector product
in Krylov-subspace algorithms, 
the sparsity of the matrix makes the computation efficient by
avoiding any zero matrix elements.

% the algorithm of 
% a matrix $H$ is shown as follows.
% The shell-model calculation satisfies these two features,
% and thus the Lanczos method is  the best solver for the shell-model
% calculations.

\begin{algorithm}
  \caption{Lanczos method}
  \label{lanczos}
  \begin{algorithmic}[1]
    \State vector $\bm{v}_1$ be an arbitrary vector with
    $||\bm{v}_1||=1$
    \State $\beta_{0}:=0$,\ \  $\bm{v}_0:=0$
    \For{$k = 1, 2, 3, \cdots $}
    \State $\bm{w} := H \bm{v}_k$
    \State $\alpha_{k}:= (\bm{v}_k \cdot \bm{w})$
    \State $T_{k,k}:= \alpha_{k}$
    \State Diagonalize $T^{(k)}$ and stop if $e_n$ converges
    \State $\bm{w} := \bm{w} - \beta_{k-1} \bm{v}_{k-1} - \alpha_k \bm{v}_k $
    \For{$l = 1, 2,  \cdots, k-2$}
    \State
    $\bm{w} := \bm{w}  - \bm{v}_l (\bm{v}_l \cdot \bm{w}) $ 
    \EndFor
    \State $\beta_k := \sqrt{ ( \bm{w}\cdot \bm{w}) }$
    \State $\bm{v}_{k+1} := \bm{w}/\beta_k$
    \State $T_{k,k+1}:=\beta_k$, \ $T_{k+1,k}:=\beta_k$
    \EndFor
  \end{algorithmic}
\end{algorithm}

The Lanczos method is one of the simplest Krylov-subspace methods
and has been considered to be the best
solver  for the LSSM.
A brief description of this method 
is shown in Algorithm \ref{lanczos}.
In the algorithm, ``$:=$'' denotes a variable assignment.
The eigenvalues of the subspace spanned by
the so-called Lanczos vectors $\bm{v}_k$
are denoted by $e_n^{(k)}$. They are called Ritz values 
and obtained by diagonalizing the tridiagonal $k$-dimension
matrix $T^{(k)}$  
\begin{equation}
  \label{eq:tmat}
  T^{(k)} = 
  \left(
    \begin{array}{ccccc}
      \alpha_1 & \beta_1  &          &             & 0 \\
      \beta_1  & \alpha_2 & \beta_2  &             &  \\
               & \beta_2  & \alpha_3 & \ddots      &  \\
               &          & \ddots   & \ddots         & \beta_{k-1} \\
      0        &          &          & \beta_{k-1} & \alpha_k
    \end{array}
  \right) .
\end{equation}
This algorithm causes a simple three-term recurrent relation:
\begin{equation}
  \label{eq:lanc-rec}
  \beta_k \bm{v}_{k+1} = H \bm{v}_{k}
  - \alpha_k \bm{v}_{k}  - \beta_{k-1} \bm{v}_{k-1} .
\end{equation}
The Ritz values approach the exact eigenvalues 
of the original matrix as $k$ increases.
The initial vector $\bm{v}_1$ can be taken arbitrarily,
{\it e.g.} random numbers 
or an approximate solution obtained in the truncated space.
The iteration in Algorithm \ref{lanczos} continues until 
the $e_n^{(k)}$ reaches convergence. 
% The vectors $\bm{v}_k$ are called the Lanczos vectors hereafter.
The Hamiltonian matrix $H$ appears only in line 4
as a matrix-vector product, 
which can be performed quite efficiently for a sparse matrix. 
% However, 
% the cost of the reorthogonalization increases as the $k$ increases.

The procedure of the lines 9, 10, and 11 in Algorithm \ref{lanczos}
is called the reorthogonalization.
In principle,  $\bm{v}_{k+1}$ is orthogonal to 
$\bm{v}_l$, $(l=1,2,\cdots,k-2)$ even without
this reorthogonalization procedure.
However, since the numerical error in actual calculations
deteriorates the orthogonality, 
this procedure is essential in practice.
Its computational cost is proportional to $k^2 D_M$ and 
the memory (or disk) capacity reaches $kD_M$, 
and it becomes a bottleneck of the computation if $k$ is large.
In order to avoid this difficulty, the thick-restart Lanczos method 
was proposed \cite{tr-lanczos},
and is discussed in the next subsection.
% In this method, the computational cost of the reorthogonalization
% is reduced by restarting the Lanczos procedure
% when the number of the Lanczos vectors reaches the upper limit $l_m$.

\subsection{Thick-restart Lanczos method }
\label{sec:trlan}

Although the Lanczos method is quite efficient,
the cost of the reorthogonalization increases and is proportional
to the number of the Lanczos vectors squared.
In order to reduce this cost, 
the thick-restart Lanczos method was proposed, 
where we restart the Lanczos iterations
by compressing the whole Lanczos vectors into a small number of
vectors having the lowest eigenvalues.
Its algorithm is shown in Algorithm \ref{tr-lanczos}.

\begin{algorithm}
    \caption{Thick-restart Lanczos method}
    \label{tr-lanczos}
    \begin{algorithmic}[1]
      \State vector $\bm{v}_1$ be an arbitrary vector with  
      $||\bm{v}_1||=1$
      \State  $k_x := 1$
      \For{$l = 1, 2, 3, ... $ }
      \For{$k = k_x, k_x+1, k_x+2, ..., l_m-1$}
      \State $\bm{w} := H \bm{v}_k$
      \State $\alpha_{k} := (\bm{v}_k \cdot \bm{w})$
      \State $T_{kk}:=\alpha_k$
      \State Diagonalize $T^{(k)}$ and stop if $e_n$ converges
      % \State $\bm{w} := \bm{w} - \beta_{k-1} \bm{v}_{k-1} - \alpha_k v_k $
      \For{$l = k, k-1, \cdots,2,1$}
      \State $\bm{w} := \bm{w}
      - \bm{v}_l (\bm{v}_l  \cdot \bm{w} ) $ 
      \EndFor
      \State $\beta_k := \sqrt{ ( \bm{w}\cdot \bm{w}) }$
      \State $\bm{v}_{k+1} := \bm{w}/\beta_k$
      \State $T_{k,k+1}:=\beta_k$, \ $T_{k+1,k}:=\beta_k$
      \EndFor
      \State Construct a new $T^{(l_s+1)}$ matrix
      and $v_{1},\cdots,v_{l_s+1}$ for restart
      \State $k_x := l_s+1$
      \EndFor
    \end{algorithmic}
\end{algorithm}

The lines 4 to 15 are the same as
the Lanczos algorithm in Algorithm \ref{lanczos}.
The outer loop represents the thick-restart procedure, 
and in line 16 we prepare the Lanczos vectors and
the $T^{(k)}$ matrix for the restart. 
In practice, the subspace spanned by the $l_m-1$ vectors
is compressed to 
that by the $l_s$ vectors
by choosing the lowest $l_s$ eigenvectors
of the subspace just before the restart.

At the restart, the $T^{(k)}$ matrix and $v$ after the restart
is constructed as
\begin{equation}
  \label{eq:tr-tmat}
  T^{(k)} := 
  \left(
    \begin{array}{cccccccc}
      e_1 &     &        & 0       & r_1            &  &  & 0 \\
          & e_2 &        &         & r_2            &  &  &  \\
          &     & \ddots &         & \vdots         &  &  &  \\
      0   &     &        & e_{l_s} & r_{l_s}        &  &  &  \\
      r_1 & r_2 & \cdots & r_{l_s} & \alpha_{l_s+1} & \beta_{l_s+1}       & & \\
          &     &        &         & \beta_{l_s+1}  & \alpha_{l_s+2}      & \ddots &  \\
          &     &        &         &                & \ddots & \ddots     & \beta_{k-1} \\
      0   &     &        &         &                &        & \beta_{k-1}& \alpha_k 
    \end{array}
  \right) , 
\end{equation}
and 
\begin{equation}
  \label{eq:tr-emat}
  E^{(l_s)} = 
  \left(
    \begin{array}{cccc}
      e_1 &     &        & 0       \\
          & e_2 &        &         \\
          &     & \ddots &         \\
      0   &     &        & e_{l_s} \\
    \end{array}
  \right) , 
\end{equation}
\begin{eqnarray}
  r_k & := & \beta_{l_m-1} U_{l_m-1,k}
  \ \ \ \textrm{for} \ \ \ k=1,2,\cdots,l_s \\
  % \alpha_{l_s+1} & = & \alpha_{l_m} \\
  \bm{v}_{k} & := &  \sum_{j=1}^{l_m-1} \bm{v}_j U_{jk} 
                    \ \ \ \textrm{for} \ \ \ k=1,2,\cdots,l_s  \\
  \bm{v}_{l_s+1} & := & \bm{v}_{l_m}
\end{eqnarray}
where $e_k$ and $U_{lk}$ are the $k$-th eigenvalue and eigenvector
of the $T^{(l_m-1)}$ matrix just before the restart.
Thus, we restart the Lanczos iterations with keeping 
the $l_s$ eigenvalues of $T^{(k)}$.
Note that the three-term recurrence is valid only after 
$l_s+2$-th vector.

The restart is done to restrict 
the number of the Lanczos vectors and its reorthogonalization costs.
After the restart, 
the $T^{(k)}$ matrix no longer keeps a tridiagonal form 
and therefore an efficient way to diagonalize the tridiagonal matrix
cannot be applied.
However, the additional computation cost to diagonalize $T^{(k)}$
is negligible since the dimension of $T^{(k)}$
is $O(10^2)$ typically and is far smaller 
than the dimension $D_M$.

Lines 9, 10 and 11 in Algorithm \ref{tr-lanczos}
contain  the orthogonalization of $\bm{w}$ 
with $\bm{v}_k$ and $\bm{v}_{k-1}$, and the reorthogonalization
with all the previous vectors $\bm{v}_1, \bm{v}_2 \cdots \bm{v}_{k-2}$.
This reorthogonalization is necessary just after
the restart even mathematically.
The performance of the reorthogonalization and its relation
to the thick restart is discussed 
in Appx.~\ref{sec:reorth}.

\subsection{Block Lanczos method }
\label{sec:blan}

In the Lanczos and thick-restart Lanczos methods, 
the matrix-vector product is a bottleneck
of the total computation time.
Especially in the KSHELL code, the matrix elements are
generated on the fly at every matrix-vector product,
namely at every Lanczos iteration.  
It costs a certain amount of the elapsed computation time.
In general, the block algorithm decreases the number of
iterations, and therefore it is expected to reduce 
the frequency of the on-the-fly generation
and consequently to shorten the elapsed time.
The idea of the block algorithm is that a certain number of 
vectors are bundled as a block 
% and the frequency of the on-the-fly generations is reduced
and the product of the matrix and the block vectors is performed at once.
The Ritz values are obtained in the subspace
spanned by the block Krylov subspace \cite{block-krylov}
\begin{eqnarray}
  \label{eq:block-krylov}
  && {\cal K}_m(H, \bm{v}_1^{(1)}, \bm{v}_1^{(2)},\cdots, \bm{v}_1^{(p)})
  \\
  &&= \{
  \bm{v}_1^{(1)}, \cdots, \bm{v}_1^{(p)}, 
     H\bm{v}_1^{(1)}, \cdots, H\bm{v}_1^{(p)},
     \nonumber \\
  && 
     \ \ \ \ \ H^2\bm{v}_1^{(1)}, \cdots, H^{m-1}\bm{v}_1^{(p)} \} ,
  \nonumber 
\end{eqnarray}
where $p$ denotes the number of the initial vectors $\bm{v}_1^{(p)}$,
or called the block size.
Hereafter, $p$ vectors are grouped as a block, or a 
$D_M\times p$ matrix
$\bm{V}_1  = \left(\bm{v}_1^{(1)}, \bm{v}_1^{(2)},\cdots, \bm{v}_1^{(p)}\right)$.
The block Krylov subspace is rewritten as
\begin{eqnarray}
  \label{eq:block-krylov-V}
  {\cal K}_m(H, \bm{V}_1)
  &=& \{
  \bm{V}_1, H\bm{V}_1, H^2\bm{V}_1, \cdots, H^{m-1}\bm{V}_1 \} .
\end{eqnarray}
As $m$ increases, the Ritz value of this subspace is expected
to converge faster than that of the Krylov subspace.

The algorithm of the block Lanczos method \cite{block-lanc}  is as follows.
In the algorithm, 
$\bm{V}_k$ and $\bm{W}$  are $D_M \times p$ matrices, and 
$\bm{\alpha}$ and  $\bm{\beta}$
are $p \times p$ matrices.

\begin{algorithm}
    \caption{Block Lanczos method}
    \label{alg:blan}
    \begin{algorithmic}[1]
      \State $\bm{V}_1$ be arbitrary vectors with  
      $\bm{V}_1^T \bm{V}_1  = \bm{1}$ 
      \For{$k = 1, 2, 3, \cdots $}
      \State $\bm{W} := H \bm{V}_k$
      \State $\bm{\alpha}_{k} := \bm{V}_k^T \bm{W}$
      \State $T_{p(k-1)+1:pk,p(k-1)+1:pk} := \bm{\alpha}_k$
      \State Diagonalize $T^{(k)}$ and stop if $e_n$ converges
      % \State $\bm{W} := \bm{W} - \bm{V}_k  \bm{\alpha}_k $
      \State Orthogonalize $\bm{W}$ with
      $ \bm{V}_1,\bm{V}_2,\cdots,\bm{V}_{k}$
      % \For{$l = k-2, k-3, \cdots, 2, 1$}
      % % \State $\bm{v}_{k+1} =
      % % \bm{v}_{k+1}  - \bm{v}_l (\bm{v}_l \cdot \bm{v}_{k+1}) $ 
      % \State $\bm{\gamma} = \bm{V}_l^T \bm{V}_{k+1}$
      % \State $\bm{V}_{k+1} = \bm{V}_{k+1} - \bm{V}_k  \bm{\gamma} $
      % \EndFor
      \State $\bm{V}_{k+1} \bm{\beta_k} :=  {\rm QR}(\bm{W})$
      \State $T_{pk+1:p(k+1), p(k-1)+1:pk} := \bm{\beta}_k$
      \State $T_{p(k-1)+1:pk, pk+1:p(k+1)} := \bm{\beta}_k^T$,
      \EndFor
    \end{algorithmic}
\end{algorithm}
In the algorithm ${\rm QR}(\bm{W})$ denotes
the QR decomposition of the matrix $\bm{W}$ \cite{num_recipe}.
``$T_{a:b, c:d}$'' denotes a submatrix of $T$
in the notation of a Fortran array section.
The line 7 in Algorithm \ref{alg:blan}
is the reorthogonalization of new vectors $W$ with all
previous Lanczos vectors,
although only the orthogonalization with $\bm{V}_k$ and $\bm{V}_{k-1}$
is enough mathematically.
This algorithm is 
similar to that of the simple Lanczos method
except that the Lanczos vectors are replaced by
the block vectors and the QR decomposition
is introduced so that the vectors of a block
are kept orthogonalized to each other.

The eigenvalues of the subspace $e_n^{(k)}$ are
obtained by diagonalizing the $pk \times pk$
symmetric block-tridiagonal matrix 
\begin{equation}
  \label{eq:tmat-blan}
  T^{(k)} = 
  \left(
    \begin{array}{ccccc}
      \bm{\alpha}_1 & \bm{\beta}_1^T  &          &             & 0 \\
      \bm{\beta}_1  & \bm{\alpha}_2 & \bm{\beta}_2^T  &             &  \\
                    & \bm{\beta}_2  & \bm{\alpha}_3 & \ddots      &  \\
                    &               & \ddots   & \ddots         & \bm{\beta}_{k-1}^T \\
      0             &               &          & \bm{\beta}_{k-1} & \bm{\alpha}_k
    \end{array}
  \right) , 
\end{equation}
which is constructed in lines 5, 9, 10 of Algorithm \ref{alg:blan}.

The block Lanczos method has two advantages over the simple Lanczos method:
one is the fact that the degenerate eigenvalues up to the block size
can be obtained accurately. 
It is helpful to obtain highly-excited states where
the level density increases and near-degeneracy would occur
in shell-model calculations,
while the simple method works more efficiently
in case of obtaining a small number of states.
The other is that, in general, a matrix-matrix product is 
far efficiently calculated than a matrix-vector product.
On the other hand, the number of the Lanczos vectors 
tends to be larger than the simple Lanczos method,
which would cause difficulty in reorthogonalization.
In order to overcome this problem, we introduce
the thick-restart method in the same way as the thick-restart
Lanczos method.

\subsection{Thick-restart block Lanczos method }
\label{sec:trblan}

% When a small number of states or the ground state is required, 
% the number of iterations is not so different between
% the Lanczos and the block Lanczos 
% methods, and under this condition, 
% the Lanczos method is more efficient than 
% the block one.

When a large number of eigenvalues are required,
the block algorithm is expected to reduce the
number of iterations and the elapsed time.
However, the number of the Lanczos vectors
tends to increase more than the simple Lanczos method 
the cost of their reorthogonalization increases accordingly.
While the implicitly restart block Lanczos method is known
to restrict the number of the Lanczos vectors \cite{ir-b-lanc},
we here propose to combine the block Lanczos method
with the thick restart to
reduce the cost of the reorthogonalization.
Its algorithm is shown in Algorithm  \ref{alg:trblan}.
Similar algorithms for a non-symmetric matrix or
a linear response eigenvalue problem 
have been discussed in
Refs.~\cite{tr-b-lanc,tr-b-arnoldi,aug-b-lanc}.

\begin{algorithm}
  \caption{Thick-restart block Lanczos method}
  \label{alg:trblan}
  \begin{algorithmic}[1]
    \State $\bm{V}_1$ be arbitrary vectors with  
    $\bm{V}_1^T \bm{V}_1  = \bm{1}$ and   $k_x:=0$.
    \For{$l = 1, 2, 3, \cdots $}
    \For{$k = 1, 2,  \cdots $}
    \State $\bm{W} := H \bm{V}_k$
    \State $\bm{\alpha}_{k} := \bm{V}_k^T \bm{W}$
    \State $T_{k_x+p(k-1)+1:k_x+pk,k_x+p(k-1)+1:k_x+pk} := \bm{\alpha}_k$
    \State Diagonalize $T^{(k)}$ and stop if $e_n$ converges
    % \State $\bm{W} = \bm{W} - \bm{V}_k  \bm{\alpha}_k $
    \State Orthogonalize $\bm{W}$ with
    $ \bm{v}_1,\bm{v}_2,\cdots,\bm{v}_{k_x+pk}$
    \State $\bm{V}_{k+1} \bm{\beta_k} :=  {\rm QR}(\bm{W})$
    \State $T_{k_x+pk+1:k_x+p(k+1),k_x+p(k-1)+1:k_x+pk} := \bm{\beta}_k$
    \State $T_{k_x+p(k-1)+1:k_x+pk,k_x+pk+1:k_x+p(k+1)} := \bm{\beta}_k^T$,
   \EndFor
   \State Construct $T^{(l_s)}$  and
   $\bm{v}_k,1\leq k \leq l_s$  for restart
   \State $k_x := l_s$
   \EndFor
 \end{algorithmic}
\end{algorithm}

The restart is done 
so that the number of Lanczos vectors does not exceed 
the given upper limit $l_m$.
The $T^{(k)}$ matrix and $\bm{v}$ after the restart is constructed
in the same way as Eq.(\ref{eq:tmat-blan}) before the restart. 
The $T^{(k)}$ matrix after the restart is constructed as 
\begin{equation}
  \label{eq:trb-tmat}
  T^{(k)} := 
  \left(
    \begin{array}{cccccc}
      E^{(l_s)}  & \bm{r}^T     &  &  & & 0 \\
      \bm{r}    & \bm{\alpha}_{1} & \bm{\beta}_{1}^T       & & & \\
                & \bm{\beta}_{1}& \bm{\alpha}_{2} & \bm{\beta}_{2}^T & & \\
                &              & \ddots & \ddots     & \ddots & \\
                &              & & \bm{\beta}_{k-2} & \bm{\alpha}_{k-1}      & \bm{\beta}_{k-1}^T \\
      0         &              & &        & \bm{\beta}_{k-1}& \bm{\alpha}_k 
    \end{array}
  \right) , 
\end{equation}
where $E^{(l_s)}$ is a diagonal matrix whose
matrix elements are the Ritz values
$(e_1, e_2, \cdots, e_{l_s})$ of the matrix $T$
which is constructed just before the restart.
While the $T^{(k)}$ matrix is no longer block tridiagonal after the restart,
it is still symmetric.
The Lanczos vectors up to $k$-th iterations after
the start or the restart are defined as
\begin{eqnarray}
  \label{eq:trblan-vecs}
  && \bm{v}_{k_x+1}, \bm{v}_{k_x+2}, \cdots ,\bm{v}_{k_x+pk}
  \\
  && :=
     \bm{v}_1^{(1)}, \bm{v}_1^{(2)},  \cdots, \bm{v}_1^{(p)},  
     \bm{v}_2^{(1)}, \cdots \cdots, \bm{v}^{(p)}_k .
  \nonumber
\end{eqnarray}
These Lanczos vectors after the restart are constructed as 
\begin{eqnarray}
  \bm{v}_{k}
  & := &  \sum_{j} \bm{v}_j U_{jk}
         \ \ \ \textrm{for} \ \ \ k=1,2,\cdots,l_s  \\
  \bm{V}_{1} & := & \bm{V}_{k_m+1} \\
  \bm{r} & := & \bm{\beta}_{k_m}
                U_{k_x+p(k_m-1)+1:k_x+pk_m, 1:l_s} 
\end{eqnarray}
where $e_k$ and $U_{lk}$ are the $k$-th eigenvalue and eigenvector
of the $T^{(k)}$ matrix before the restart.
The $k_m$ denotes the $k$ just before the restart.
Note that $l_s$ is not necessarily a multiple of $p$.

Thus, the thick-restart procedure again 
enables us to restrict the number of the Lanczos vectors
and to reduce the cost of the reorthogonalization,
which tends to increase in the block  algorithm.
% For the compensation, this method enables 
% us to make both the matrix-vector operation
% and reorthogonalization efficiently.

\section{Performance of the Lanczos methods}
\label{sec:perf}

In the previous section, we briefly introduced
the four methods of the solver for the
eigenvalue problem:
the simple Lanczos, the thick-restart Lanczos,
the block Lanczos, and the thick-restart block Lanczos
methods.
The convergence properties of these four Lanczos 
methods are discussed in Sect.~\ref{sec:conv}.
Their performance
is compared in Sect.~\ref{sec:perf-block}.

\subsection{Convergence of the Lanczos method and its variants}
\label{sec:conv}

The convergence properties
of the four Lanczos methods in the LSSM calculations
are discussed in this subsection.
We take $^{48}$Cr with the $pf$-shell model space and
the GXPF1A interaction \cite{gxpf1a}
as an example throughout this subsection.
In this case, 4 protons and 4 neutrons occupy
$pf$-shell orbits, which consist of
$0f_{7/2}, 0f_{5/2}, 1p_{3/2},$ and $1p_{1/2}$ single-particle orbits,
or the 20 single-particle states both for protons and neutrons. 
Its $M$-scheme dimension is 1,963,461.

\begin{figure}[htbp]
%  \begin{center}
  \includegraphics[scale=0.4]{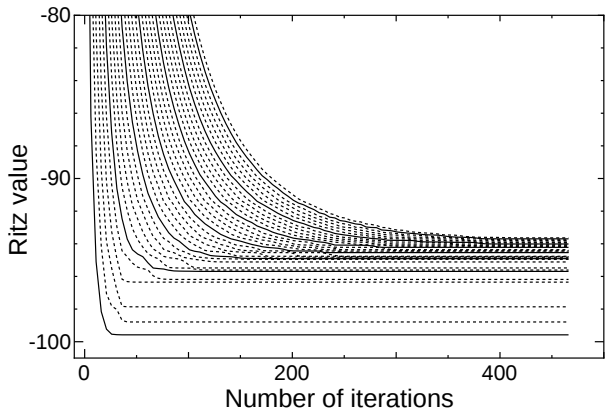}
%  \end{center}
  \caption{
    Convergence of the simple Lanczos method
    in the case of $^{48}$Cr with
    the GXPF1A interaction.
    The lines denote  the lowest 32 Ritz values
    against the number of Lanczos iterations.
    The 1st, 6th, 11th, 16th, 21st, 26th,
    and 31st Ritz values
    are indicated by the solid lines,
    while the dotted lines denote the other values.
  }
  \label{fig:lanczos-cr48}
\end{figure}

Figure \ref{fig:lanczos-cr48} shows
the convergence of the 32 eigenvalues 
as a function of the number of Lanczos iterations.
The criterion of convergence is that  
the change of the Ritz values as a function of the number
of iterations is smaller than $10^{-6}$ MeV,
which is small enough for practical usage.
The lowest eigenvalue converges quite fast 
and reaches the convergence at the 44th iteration.
On the other hand, higher eigenvalues converge slower, 
and the 32nd one reaches convergence at the 466th iteration.
In this case, the whole 467 Lanczos vectors should be stored
for the reorthogonalization
and for obtaining the eigenvectors if necessary.

\begin{figure}[htbp]
  \includegraphics[scale=0.4]{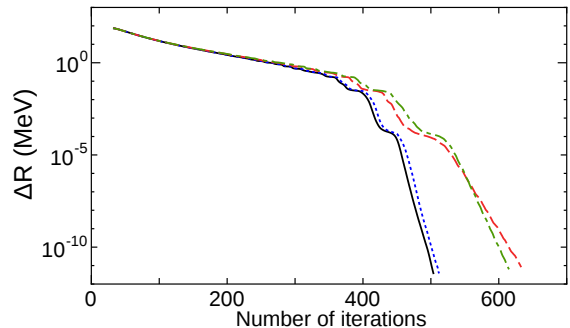}
  \caption{
    Convergence of the Lanczos
    and thick-restart Lanczos methods
    in the case of the 32nd lowest eigenvalue of
    $^{48}$Cr with the GXPF1A interaction.
    The deviation of the 32nd lowest Ritz value from
    the exact eigenvalue  ($\Delta R$)
    is shown against the number of Lanczos iterations.
    The solid black line is provided by the Lanczos method,
    while the convergence of the thick-restart Lanczos method
    is denoted by the blue dotted ($l_s=40, l_m=100$),
    green dot-dashed ($l_s=40, l_m=50$),
    and red dashed ($l_s=36, l_m=50$)  lines.
  }
  \label{fig:lanczos-32nd}
\end{figure}

While the thick-restart method can be used to reduce
the reorthogonalization cost,
the frequent restarts may deteriorate the convergence. 
We discuss the convergence of the thick-restart Lanczos method
in Fig. \ref{fig:lanczos-32nd}.
In the figure the deviation between the 32nd lowest Ritz value
and the exact eigenvalue is shown.
Hereafter, we focus on the convergence only
of the 32nd eigenvalue without particular mention.
The number of iterations
means the number of accumulated Lanczos steps,
not the number of restarts.
The thick-restart Lanczos method with $l_s=40$ and $l_m=100$
shows reasonably fast convergence and
requires modest storage (100 Lanczos vectors),
while small $l_m$ ($l_m=50$) deteriorates 
the convergence
since the restart occurs too frequently.
The small $l_s$ ($l_s=36$) also deteriorates
the convergence 
due to the loss of the components at the restart.

\begin{figure}[htbp]
  \includegraphics[scale=0.4]{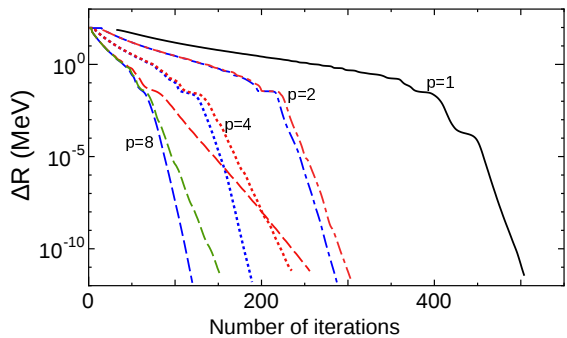}
  \caption{
    Convergence of the block Lanczos
    and thick-restart block Lanczos methods.
    The deviation of the 32nd lowest Ritz value
    from the exact eigenvalue ($\Delta R$)
    is shown against the number
    of iterations.
    The solid line denotes the convergence of
    the simple Lanczos method.
    The convergences of the block Lanczos method
    are denoted by the blue dot-dashed ($p=2$),
    dotted ($p=4$), and dashed ($p=8$) lines, respectively, 
    with $p$ being the block size.
    The corresponding results of
    the thick-restart block Lanczos method
    to restrict the storage size up to the 100 Lanczos vectors
    ($l_m=100$)  are shown by the red lines.
    The green dashed line denotes the $p=8$
    convergence with  $l_m=200$ and $l_s=40$.
    See the caption of Fig.~\ref{fig:lanczos-32nd} for
    further details.
  }
  \label{fig:block-lanczos-32nd}
\end{figure}

Figure \ref{fig:block-lanczos-32nd} shows the
convergence of the Ritz values by
the block Lanczos and thick-restart block Lanczos
methods.
In the block method, the number of iterations
is equal to the number of products of the matrix and block vectors.
While the $p=1$ line shows the results of the simple Lanczos method,
those of the block Lanczos method with the block sizes
$p=2, 4,$ and 8 are shown as the blue lines and
reach convergence in a small number of iterations.
For the block method, the number of iterations for convergence 
is almost equal and slightly larger than $1/p$ of the simple Lanczos method.
It means that the total number of the Lanczos vectors
needed in the block method 
is slightly larger than that of the simple Lanczos method. 
This additional cost is overwhelmed by
the acceleration of the product of the matrix and a block of vectors, 
the details of which are discussed in the next subsection.

The restart algorithm enables us to
decrease the number of Lanczos vectors
to be stored, and the cost of the reorthogonalization.
The red line in Fig.~\ref{fig:block-lanczos-32nd}
shows the convergence of the
thick-restart block Lanczos method
with the number of vectors restricted up to 100 ($l_m=100$). 
While the convergence becomes slightly slow in comparison with the block Lanczos
method in the case of $p=2$ and $p=4$,
this additional cost is compensated 
by the speedup of the reorthogonalization.
However, in the case of $p=8$ with $l_m=100$, 
the convergence is quite slow since
the restart occurs too frequently.
The $p=8$ case with the $l_m=200$ 
reduces the number of the iterations and
it approaches that of the block Lanczos method without restart.

\subsection{Performance of the block algorithm}
\label{sec:perf-block}

In the previous section, we showed that the block algorithm
decreases the number of iterations drastically.
However, since the elapsed time of a product of a
matrix and a block of the vectors increases with $p$,
the total performance depends on the balance of the number
of iterations and the increased cost of the matrix-block product.
In this subsection, we describe
some examples of this trade-off.
Besides, the thick-restart algorithm also
causes a trade-off between the reorthogonalization cost
and the number of iterations.
The detail of the latter trade-off is discussed in
Appx.~\ref{sec:reorth}.

In the KSHELL code, in order to save the memory size 
the matrix-vector product is realized 
by the on-the-fly generation of the matrix elements
discussed in Appx.~\ref{sec:matvec}.
In order to reduce the additional cost of this generation
we adopt the block algorithm. 
In the usual Lanczos method, the matrix elements
are generated on the fly at every matrix-vector product.
On the other hand, in the block algorithm, the generation is 
done once for a bundle of vectors, which are taken 
as a block.
Thus, the block algorithm is expected to reduce the total
elapsed time.

%% \begin{figure}[htbp]
%%   \includegraphics[scale=0.4]{block-matvec-time-cr48.eps}
%%   \caption{
%%     Elapsed time of the products of the matrix and the block vectors.
%%     The red circles denote 
%%     the time of a product of a matrix and a block vectors
%%     against the block size $p$. 
%%     The blue triangles denote the elapsed time per vector.
%%     The dotted lines are drawn to guide the eyes.
%%   }
%%   \label{fig:time-block-matvec}
%% \end{figure}

\begin{figure}[htbp]
  \includegraphics[scale=0.4]{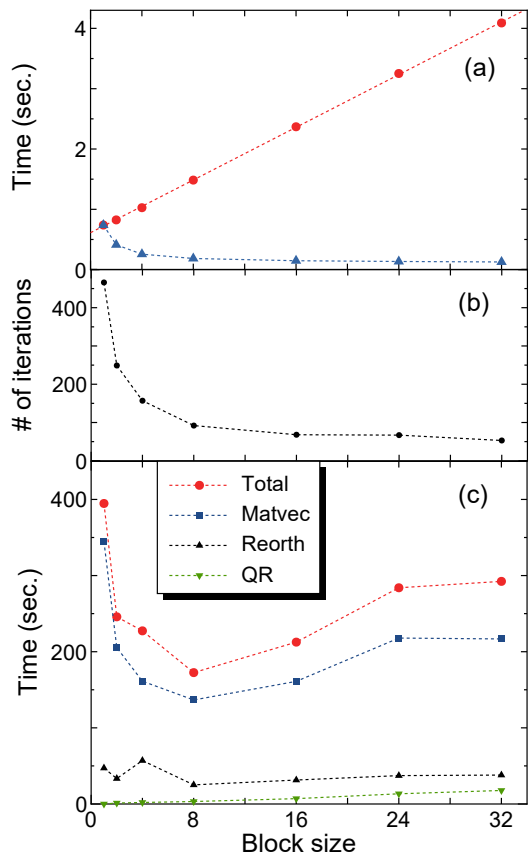}
  \caption{
    Performance of the thick-restart block Lanczos method
    to obtain the 32 lowest eigenvalues 
    of $^{48}$Cr with the GXPF1A interaction 
    with the KSHELL code. 
    (a) Elapsed time of a product
    of the matrix and the block vectors.
    The red circles denote 
    the time of a product of a matrix and a block vectors
    against the block size $p$. 
    The blue triangles denote the elapsed time per vector.
    The dotted lines are drawn to guide the eyes.    
    (b) The number of iterations for convergence. 
    (c) The red circles, blue squares, black triangles,
    and green inverted triangles denote
    the times for total computation, products of the matrix and
    block vectors,
    reorthogonalization, and QR-decomposition, respectively.
  }
  \label{fig:time-total-blan}
\end{figure}

Figure \ref{fig:time-total-blan} (a) shows
the elapsed time of a product of the matrix
and a block of vectors 
in the LSSM of $^{48}$Cr, which was also
discussed in the previous subsection.
% In this case, 4 protons and 4 neutrons occupy
% $pf$-shell orbits and 
% the same configuration spaces of protons and neutrons, 
% and the factorization algorithm discussed
% in sect.~\ref{sec:factpn}
% works efficiently to shorten the time of the on-the-fly generation
% of the matrix elements.
The performance 
was measured by the KSHELL code on 20 CPU cores of Intel Xeon E5-2680.
All the Lanczos vectors are stored on memory.
In the case of $p=1$ corresponding to the simple Lanczos method,
one matrix-vector product costs 0.73 sec.
By increasing $p$, the time increases and can be fitted by a line.
The $y$-intercept of the fitted line, 0.6 seconds,
is the overhead cost of the on-the-fly generation of the 
matrix elements.
This overhead cost is fixed and rather
independent of $p$.
Therefore, as $p$ increases this overhead cost
becomes negligible relative to the time of a matrix product per vector
(blue triangle in Fig.\ref{fig:time-total-blan} (a)).
As $p$ increases, the time per vector approaches 0.13 seconds, 
which corresponds to the gradient of the red fitted line.

Figure \ref{fig:time-total-blan} (b) shows
the number of iterations of the thick-restart block Lanczos method 
to obtain the lowest 32 eigenvalues 
of $^{48}$Cr.
It decreases drastically as a function of the block size $p$,
and reaches 53 at $p=32$, which is almost 1/9 smaller than the case of
$p=1$, 466.

Figure \ref{fig:time-total-blan} (c) shows
the total elapsed time of the LSSM of the $^{48}$Cr case
as a function of $p$.
The $p=8$ case shows the shortest time,
which is determined 
by the trade-off between the number of iterations
and the time of a product of the matrix and a block of vectors.
The total time of the products of the matrix and block vectors
is shown in the figure and occupies roughly 80\% of
the total elapsed time.

\begin{figure}[htbp]
  \includegraphics[scale=0.4]{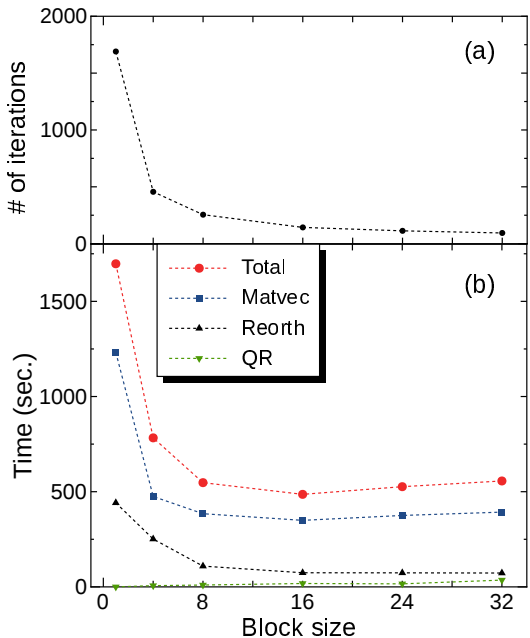}
  \caption{
    Performance to obtain the 128 lowest eigenvalues 
    of $^{48}$Cr 
    using the thick-restart block Lanczos method
    with $l_m=800$.
    See caption of Figs.~\ref{fig:time-total-blan} (b) and (c)
    for details.
  }
  \label{fig:time-ne128}
\end{figure}

The acceleration caused by
the thick-restart block Lanczos method
is more effective 
when a larger number of the eigenvalues are computed. 
Figure \ref{fig:time-ne128} shows
the elapsed time to obtain
the 128 lowest eigenvalues utilizing the thick-restart
block Lanczos method with $l_m=800$.
The other conditions are
the same as Fig.~\ref{fig:time-total-blan}.
The $p=16$ case reaches the shortest time
and provides us with 3.5 times speedup in comparison
with the thick-restart Lanczos method.
Without the thick restart, the number
of the Lanczos vectors increases to 2240
for $p=8$ and the cost of the reorthogonalization
extends the total elapsed time by 20\%
when the whole Lanczos vectors are stored on memory.

We also performed a benchmark test of the system without valence protons,
which means that the proton-neutron factorization
in Sect.~\ref{sec:factpn} does not work. 
In such case the cost of the on-the-fly matrix-elements generation
is dominant over the total elapsed time
and the block algorithm is advantageous.
As such an example we take  $^{112}$Sn
with the $50\leq N \leq 82$ model
space, namely 12 active neutrons in the
$0g_{7/2}, 1d_{5/2}, 1d_{3/2}, 2s_{1/2},$ and
$0h_{11/2}$ single-particle orbits.
Its $M$-scheme dimension is 6,210,638.
The SNBG3 interaction is adopted \cite{snbg3,La133-palit}.

%% \begin{figure}[htbp]
%%   \includegraphics[scale=0.38]{sn112-matblock-time.eps}
%%   \caption{
%%     Elapsed time of the products of the matrix
%%     and the block vectors of $^{112}$Sn
%%     as a function of the block size.  
%%   }
%%   \label{fig:sn112-time-block}
%% \end{figure}

\begin{figure}[htbp]
  \includegraphics[scale=0.4]{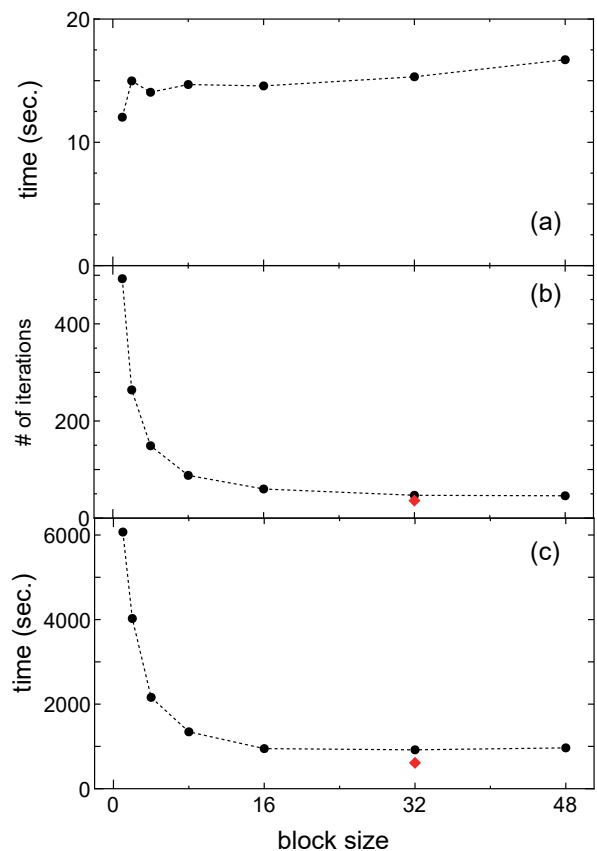}
  \caption{
    Performance of the thick-restart block Lanczos method
    to obtain the 32 lowest states of 
    $^{112}$Sn with the SNBG3 interaction \cite{snbg3}.
    The computation was performed at a single node of 
    Oakforest-PACS computer.
    (a) Elapsed time of the products of the matrix
    and the block vectors of $^{112}$Sn
    as a function of the block size.      
    The number of the Lanczos iterations (b)
    and the elapsed time (c) are shown as the black circles.
    The red diamond shows the 
    best performance by utilizing 
    the initial vectors which are prepared by the particle-hole
    truncated approximation.
    See text for details. }
  \label{fig:block-perf-sn112}
\end{figure}

Figure \ref{fig:block-perf-sn112} (a)  shows the elapsed times
of a product of the Hamiltonian matrix and a block of vectors.
It was performed on a single node
of the Oakforest-PACS computer 
equipped with 68 CPU cores of Intel Xeon Phi 7250 \cite{oakforest-pacs}.
The code runs with 272 threads for hyperthreading.
Unlike the case of $^{48}$Cr, the elapsed time shows
small dependence on the size of the block
since it is dominated by the cost of the on-the-fly generation.
Therefore the acceleration of the block algorithm
is expected to increase in this case.
Ideally the relation between the time and the block size
should be linear, but fluctuations are seen possibly because of the
cache-related matter.

Figure \ref{fig:block-perf-sn112} (b) shows
the number of iterations 
to reach the convergence of the 32 lowest eigenvalues
as a function of the block size $p$.
The number of the iterations drastically
decreases as $p$ increases, and it reaches the
smallest one, 47, at $p=32$.
As the number of iterations decreases
the elapsed time also decreases drastically
as shown in Fig.~\ref{fig:block-perf-sn112} (c).
% The $p=32$ case shows the shortest time
% o compute the 32 lowest eigenvalues.

In these benchmarks so far 
the elements of the initial vectors are taken randomly.
On the other hand, well-approximated wave functions 
can also be used as initial vectors
and are expected to accelerate 
the convergence in the block method.
As a benchmark test, we prepare 32 initial vectors 
by diagonalizing the Hamiltonian 
in the truncated subspace up to 4-particle 4-hole excitation
across the $N=64$ subshell gap. The computation time with
the truncated subspace is negligibly small.
The red diamonds in Fig.~\ref{fig:block-perf-sn112}
show the best case utilizing those well-approximated
initial vectors.
As a consequence, this best case takes 672 seconds which is
much accelerated in comparison with the case without block algorithm, 
6,077 seconds. 
Note that such a remedy cannot be applied to the simple Lanczos method
since the Lanczos method can use only one initial guess.

\section{Summary}
\label{sec:summary}

We introduced the thick-restart block Lanczos method
as an eigensolver for large-scale shell-model calculations
and discussed its performance in comparison
with the conventional Lanczos method.
Especially when a large number of eigenvalues
are required, the block method drastically reduces
the number of iterations
and the additional cost of 
the on-the-fly generation of the matrix elements
in the KSHELL code. 
Moreover, the thick-restart algorithm
restricts the number of Lanczos vectors
and reduces the cost of the reorthogonalization.

The $M$-scheme shell-model code KSHELL was developed 
for massively parallel computation and
is advantageous to obtain highly excited states
thanks to the thick-restart block Lanczos method. 
We demonstrated that the thick-restart block Lanczos method
succeeds in reducing the elapsed time of the LSSM calculations
by utilizing the KSHELL code taking $^{48}$Cr and $^{112}$Sn
as examples.
The performance of the KSHELL code is further
discussed in the appendices. 

%% Generally, the memory capacity is the strictest restriction 
%% to perform large-scale calculations
%% for $M$-scheme shell model codes.
%% Although the Hamiltonian matrix is sparse,  
%% the number of the non-zero matrix elements are 
%% too huge to be stored on memory.
%% Our strategy is to construct the matrix elements 
%% for every matrix-vector product on-the-fly
%% to save the memory usage.
%% It enables us to perform The feasibility of the $M$-scheme dimension
%% reaches $10^{11}$. 

It would be interesting to
discuss the nuclear finite-temperature properties using
the Lanczos methods \cite{ft-lanczos}.
Pursuing the possibility of the block algorithm, 
the block Sakurai-Sugiura method 
using the z-Pares package \cite{block-ss, z-pares}
provides us with promising results,
which will be reported in another publication.

\begin{acknowledgments}
  
  This work was partly supported by KAKENHI grants 
  (17K05433, 25870168, 15K05094)
  from JSPS, 
  the HPCI Strategic Program Field 5, 
  Priority issue 9 
  to be tackled by using Post K Computer from MEXT and JICFuS, 
  and the CNS-RIKEN joint project 
  for large-scale nuclear structure calculations. 
  The numerical calculation was performed partly
  on the FX10 supercomputer at the University of Tokyo, 
  K computer at AICS (hp170230, hp180179), Oakforest-PACS for
  Multidisciplinary Computational Sciences Project 
  of Tsukuba University (xg18i035).

  NS acknowledges T. Abe, Y. Futamura, M. Honma,
  T. Ichikawa,  Y. Iwata, C. W. Johnson, H. Matsufuru,
  T. Miyagi, J. E. Midtb\o, T. Otsuka, 
  C. Qi, T. Sakurai, T. Togashi, N. Tsunoda, 
  S. Yoshida, T. Yoshida, and C. Yuan  for valuable discussions,
  contributions and/or tests of the KSHELL code.
  
\end{acknowledgments}

\appendix

\section{$M$-scheme basis states and partitions}
\label{sec:part}

% In shell-model calculations, 
% we assume that a nucleus is composed 
% of an inert core and active particles 
% that move in some active orbitals. 
% The active particles and active orbitals 
% are usually taken as valence particles and the orbitals in the valence shell,
% respectively.
% A set of the active ones is called the model space.
% We treat the many-body correlations fully beyond mean-field approach
% within the model space.
% 
%% This approach is called the shell-model calculation, 
%% or specifically large-scale shell-model (LSSM) calculation. 
%% Especially in recent {\it ab initio} 
%% no-core shell-model calculations, in which an inert core is not assumed, 
%% it is also called as configuration interaction (CI) calculation
%% like in quantum chemistry. 
% In this paper, we concentrate on rather conventional 
% shell-model calculations with assuming an inert core,
% but almost all the contents are also
% applicable to no-core shell model calculations.
The code development plays 
a key role to develop a frontier of the LSSM calculations.
In the last two decades,
more than a dozen of shell-model codes had been developed
(e.g. ANTOINE \cite{antoine}, BIGSTICK \cite{bigstick}, 
EICODE \cite{eicode}, KSHELL \cite{kshell},
NATHAN \cite{nathan}, NuSHELL \cite{nushell,nushellx},
MFDn \cite{MFDn,mfdn-lobpcg},
MSHELL64 \cite{mshell,mshell64}, OXBASH \cite{oxbash},
and VECSSE \cite{vecsse}), 
while their algorithms are different with each other in details.
%% Nevertheless, the applicable region 
%% of the shell-model studies is 
%% restricted mainly to the $sd$-shell and $pf$-shell regions
%% and a heavier-mass region near semi-magic nuclei. 
%% It is desirable to extend the applicability to heavy-mass region.

In the present appendix, 
we describe the 
$M$-scheme basis states and how to treat them in the KSHELL code, 
which was written from scratch in Fortran 95 and
Python version 2.6
and is applicable for massively parallel computation.
For parallel computation, the $M$-scheme basis
states are divided into small groups classified by
the number of occupations and
the $z$-component of the proton angular momentum. 
This group is called ``partition'' and will be discussed later.
% is discussed in the latter half of the present appendix.

The $M$-scheme basis state is defined in Eq.~(\ref{eq:mbasis}).
A set of the occupied state
$(a_{i,1}, a_{i,2}, a_{i,3}, \cdots , a_{i,A}) $
is expressed numerically 
with occupied and unoccupied states being the bit 1 and bit 0
on the KSHELL code. % for practical computations.

\begin{table}[htb]
  \centering
  \begin{tabular}{l|cccccc|cc|c}
   \ \ \ \ \ \  $j$ & $d_{5/2}$  &&&&&& $s_{1/2}$ &&\\ 
    \#  \textbackslash $m$   & -5/2 & -3/2 & -1/2 & 1/2 & 3/2 & 5/2 & -1/2 & 1/2 & $\textbf{N}^{(p)}$ \\ \hline \hline
    1     &   1  &   0  &  0  &   1  &  0  &  1  &   0  &  0 &  (3,0) \\ 
    2     &   0  &   1  &  1  &   0  &  0  &  1  &   0  &  0 &  \\ 
    3     &   0  &   1  &  0  &   1  &  1  &  0  &   0  &  0 &  \\ \hline
    4     &   1  &   0  &  0  &   0  &  0  &  1  &   0  &  1 & (2,1)\\ 
    5     &   0  &   1  &  0  &   0  &  1  &  0  &   0  &  1 & \\ 
    6     &   0  &   1  &  0  &   0  &  0  &  1  &   1  &  0 & \\ 
    7     &   0  &   0  &  1  &   1  &  0  &  0  &   0  &  1 & \\ 
    8     &   0  &   0  &  1  &   0  &  1  &  0  &   1  &  0 & \\ \hline
    9     &   0  &   0  &  0  &   1  &  0  &  0  &   1  &  1 & (1,2)
  \end{tabular}
  \caption{
    Bit representation of the $M$-scheme basis states.
    It represents all the basis states of three particles
    occupying the $d_{5/2}$ and $s_{1/2}$ orbits
    having total $z$-component of the angular momentum $M=1/2$. 
    ``1'' and ``0''  denote occupied and unoccupied states, respectively.
    The leftmost column denotes a serial number of each basis state. 
    The rightmost column shows the occupation numbers
    of the $d_{5/2}$ and $s_{1/2}$ orbits.
  }
  \label{tab:bits}
\end{table}

As an example, we show a bit representation
of the system in which three identical particles occupy
the $d_{5/2}$ and $s_{1/2}$ single-particle orbits
in Table \ref{tab:bits}.
The $d_{5/2}$ ($s_{1/2}$) orbit consists of 
the $m=-\frac52, -\frac32, -\frac12, \frac12, \frac32$ and $\frac52$
($m=-\frac12$ and $\frac12$)
single-particle states
% and the $s_{1/2}$ consists of the  $m=-\frac12$ and $\frac12$ single-particle states
where $m$ denotes the $z$-component of the angular momentum
of the single-particle state.
The table shows the 9 Slater determinants having total $M=\frac12$.
Each $M$-scheme Slater determinant is
expressed as a binary number of which 0 and 1 denote
the unoccupied and occupied states, respectively.
These 9 determinants are divided into three groups,
$\textbf{N}^{(p)}$=(3,0), (2,1), and (1,2),
which are the occupation numbers of the $d_{5/2}$ and the $s_{1/2}$ orbits.
Each Slater determinant is labeled by a serial number, which
is shown in the leftmost column of Table \ref{tab:bits}.
In the practical algorithm, we generate and store all binary numbers 
representing the proton Slater determinants
and the neutron Slater determinants.
Any Slater determinant is represented as 
a product of the proton and the neutron Slater determinants.

In order to apply an arbitrary truncation scheme 
and to perform parallel computations efficiently, 
we split whole the $M$-scheme space into small partitions
by specifying the occupation numbers
of each single-particle orbit and the $z$-component
of angular momentum of protons $M^\pi$. 
The practical computation is performed in units of the partitions.
The occupation numbers of proton single-particle orbits and
neutron single-particle orbits 
are written as $\textbf{N}^{(p)} = (N^{(p)}_1, N^{(p)}_2, ... )$
and $\textbf{N}^{(n)} = (N^{(n)}_1, N^{(n)}_2, ... )$, 
where the subscript denotes the index of the single-particle orbits
in the model space.
A partition is specified by 
$\Gamma = (\textbf{N}^{(p)} , \textbf{N}^{(n)} , M^{(p)} )$.
Note that the parity quantum number
is specified by $\textbf{N}^{(p)}$ 
and $\textbf{N}^{(n)}$ uniquely.
% only the partitions which conserve the total parity are allowed.

\begin{figure}[htbp]
  \includegraphics[scale=0.4]{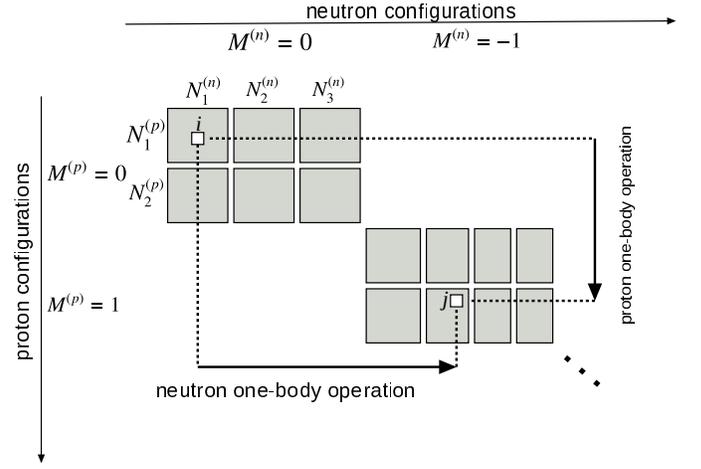}
  \caption{Conceptual drawing of the partitions
    and the proton-neutron factorization of a vector $v_i$
    with $M^{(p)}+M^{(n)}=0$.
    Each shaded box denotes a partition specified by
    $(\textbf{N}^{(p)},\textbf{N}^{(n)},M^{(p)})$.
    See text for further details.}
  \label{fig:part}
\end{figure}

The $M$-scheme subspace in a partition is written as a product of 
the proton $M$-scheme space $(\textbf{N}^{(p)}, M^{(p)})$
and the neutron $M$-scheme space, 
$(\textbf{N}^{(n)}, M^{(n)}=M^\textrm{tot}-M^{(p)})$, as
\begin{equation}
  |M_{i}\rangle = |M^{(p)}_{p_i} \rangle \otimes |M^{(n)}_{n_i} \rangle,
  \label{eq:part}
\end{equation}
where $p_i$ and $n_i$ are the indices 
of a proton partition $(\textbf{N}^{(p)}, M^{(p)})$ and 
a neutron partition $(\textbf{N}^{(n)}, M^\textrm{tot}-M^{(p)})$, respectively.
The $M$-scheme basis index $i$ is labeled by
$( \textbf{N}^{(p)}, \textbf{N}^{(n)}, M^{(p)}, p_i, n_i)$.
Figure~\ref{fig:part} shows a schematic view of the
$M$-scheme vector concerning partitions.
A vector in the $M=0$ subspace 
is split into the partitions indicated by the shaded boxes in the figures. 
Each block is specified by $ ( \textbf{N}^{(p)}, \textbf{N}^{(n)}, M^{(p)})$.

In the case of nuclear shell-model calculations, 
the Hamiltonian matrix $H_{ij}$ is very sparse 
since the Hamiltonian consists only
of one-body and two-body interactions
and the matrix element between two Slater determinants in which
more than two particles occupy different states 
is always zero.
Especially in medium-heavy nuclei,
the $M$-scheme dimension of the Hamiltonian matrix is often 
quite huge but sparse. 
Since the matrix is quite
sparse and only a few low-lying eigenstates are needed
in many LSSM calculations, 
the Lanczos method has been widely used.
% The Lanczos algorithm, one of the most famous Krylov subspace algorithms, 
% was introduced in the 1970s, 
% \cite{lanczos, lanczos-sebe, M-lanczos}
% and has been widely used in shell-model calculations. 
% Nowadays, it is implemented to take advantage of 
% massively parallel computations \cite{MFDn,bigstick}.

% Nevertheless, the application is still hampered
% by the exponential growth of the dimension of the Hilbert space. 
% The size of a Lanczos vector often surpasses the memory capacity 
% and the truncation of the model space is required.
% One of the most naive truncation methods is to assume a shell gap 
% for single-particle occupation and to restrict the number of 
% particle-hole excitation across the gap up to $t$ particles.
% It is called the $t$-particle $t$-hole truncation 
% and frequently used in practical calculations \cite{caurier_rmp}.
% As $t$ increases, the eigenstate in the truncated subspace 
% approaches the true eigenstate rather gradually.
% However, the significance of the large $t$ component remains
% and is difficult to be estimated
% (e.g. \cite{horoi_ni56,mizusaki-extrap}).
% Another frequently-used truncation scheme is
% to restrict the model space by the number of the harmonic oscillator quanta,
% or the $N_{\textrm{max}}\hbar\omega$ truncation,
% which is advantageous to remove the contamination of
% the center-of-mass motion and 
% is frequently used in no-core shell model approach
% \cite{ncsm}.
% These two truncations can be realized flexibly  
% by the selections of the partitions for the model space. 

In the $M$-scheme code, we solve the eigenvalue problem
in the subspace having good quantum numbers, 
the $z$-component of total angular momentum $M$
and the parity $\pi$.
In addition, the shell-model Hamiltonian 
has the other symmetry, namely, total angular momentum squared, $J^2$.
The resultant eigenvector becomes the eigenvector
of $J^2$ after convergence.
When only states having a specified eigenvalue
of $J^2$ are required,
at every Lanczos iteration
we can project out the Lanczos vector 
to the good $J^2$ subspace by the Lanczos diagonalization of $J^2$
\cite{caurier_rmp,mizusaki-ss}.

\section{On-the-fly algorithm of the matrix-vector product}
\label{sec:matvec}
% 
% precondition : 
%  matrix ... on-the-fly generation
%  vector ... huge, communication cost
%  matrix-product ... diagonal heavy, important for load-balancing
%

The matrix-vector multiplication is the most
time-consuming operation in the LSSM calculations.
The Hamiltonian matrix is very sparse but 
requires a huge size of memory if the whole matrix is stored.
In order to store the whole 
Hamiltonian matrix, e.g.,  in the case of $^{56}$Ni in the $pf$ shell, 
whose $M$-scheme dimension is 1,087,455,228, 
it is required to store $1.2\times 10^{12}$ non-zero matrix elements
\cite{bigstick},
namely, 14.4 TB storage in the compressed sparse raw format
and it is impractical. 
In the KSHELL code, in order 
to avoid storing the Hamiltonian matrix explicitly
we adopt the ``on-the-fly'' algorithm whose basic idea was suggested
in the 1970s and 
has been used till now \cite{M-lanczos,vecsse,caurier_rmp}.
Such on-the-fly method is utilized 
in several codes, such as ANTOINE code \cite{antoine} and MSHELL64 \cite{mshell64}.
In this section, we briefly describe how to implement the matrix-vector product 
without storing the Hamiltonian matrix elements.
In Sect.~\ref{sec:ident}, the algorithm of the one-body and two-body interactions 
between identical particles are described. 
The technique to accelerate the matrix-vector product
concerning the proton-neutron interactions 
utilizing the factorization of the proton and neutron subspaces is discussed 
in Sect.~\ref{sec:factpn}. 
This idea of the factorization was
further developed in the BIGSTICK code 
\cite{bigstick}.

\subsection{Interaction between identical particles }
\label{sec:ident}
In this appendix, we describe how the matrix-vector product is implemented 
in the case of proton-proton or neutron-neutron interactions. 
Each $M$-scheme Slater determinant $|M_i\rangle$ in Eq.(\ref{eq:mbasis}) 
is identified as a binary number:  
an occupied single-particle state of $|M_i\rangle$ 
is presented as ``bit 1'', and an unoccupied state is ``bit 0''.

The operation of the two-body interaction is performed
by bitwise operations. 
For example, the operation of a two-body term on the
$i$-th vector element is
\begin{equation}
  \label{eq:twobodybit}
  v'_j |M_j \rangle = h^{(2)}_{abcd}
  c^\dagger_a c^\dagger_b c_d c_c v_i  | M_i \rangle,
\end{equation}
where the 
binary number of $M_j$ is obtained
as the bit creation of $a$-th and $b$-th bits 
and the bit annihilation of $c$-th and $d$-th bits. 
When $a,b,c,d$ and $M_i$ are given, we obtain the binary number of $M_j$ 
and subsequently its serial number, $j$,
is determined by the binary search to find the $M_j$ in the table
such as Tab.~\ref{tab:bits}, 
which is one of the most time-consuming parts. 
The $v'_j$ is obtained as $v'_j = s h^{(2)}_{abcd} v_i $
with $s$ being a sign due to 
the anti-commutation relation.
The case of one-body interaction is obtained in the same way
straightforwardly.

\subsection{Factorization of the proton and neutron spaces}
\label{sec:factpn}

The $M$-scheme model space is spanned by
the summation of the partitions,
each of which is constructed as a
factorization of the proton and neutron subspaces 
such as Eq. (\ref{eq:part}).
The conceptual drawing of the vector $v_i$ is
shown in Fig. \ref{fig:part}.

The operation of the two-body proton-neutron interaction on the
$i$-th vector element is
\begin{eqnarray}
  \label{eq:pntwobodybit}
  v'_j |M_j \rangle
  &=& h_{abcd}
      c^\dagger_a c^\dagger_b c_d c_c v_i  | M_i \rangle,
      \nonumber \\                        
  &=& h_{abcd} v_i 
  c^\dagger_a c_c | M^{(p)}_{p_i} \rangle \  c^\dagger_b c_d  | M^{(n)}_{n_i} \rangle
      \nonumber \\                        
  &=& h_{abcd} v_i 
      | M^{(p)}_{p_j} \rangle \  | M^{(n)}_{n_j} \rangle
\end{eqnarray}
where $a$ and $c$ ($b$ and $d$) denote 
single-particle states of protons (neutrons).
Thus, the operation of the proton-neutron interaction causes the product of 
proton one-body and neutron one-body operations.

In practical computation,
we calculate and store all proton (neutron)
one-body operation 
$| M^{(p)}_{p_j} \rangle = c^\dagger_a c_c  | M^{(p)}_{p_i} \rangle $
($| M^{(n)}_{n_j} \rangle = c^\dagger_b c_d  | M^{(n)}_{n_i} \rangle $)
for every partition.
These proton and neutron one-body operations are called
``one-body jumps'' in Ref.~\cite{bigstick}.
By using these one-body jumps, we perform the summation
in Eq. (\ref{eq:twobodybit}) 
so that we avoid computing the two-body jump
in each basis. 

Figure \ref{fig:part} shows the conceptual drawing
of the factorization algorithm.
The proton and neutron one-body operations
are denoted as the solid arrows.

\section{Parallel computation and its performance}
\label{sec:perf-parallel}

\begin{figure}[htbp]
%  \begin{center}
  \includegraphics[scale=0.4]{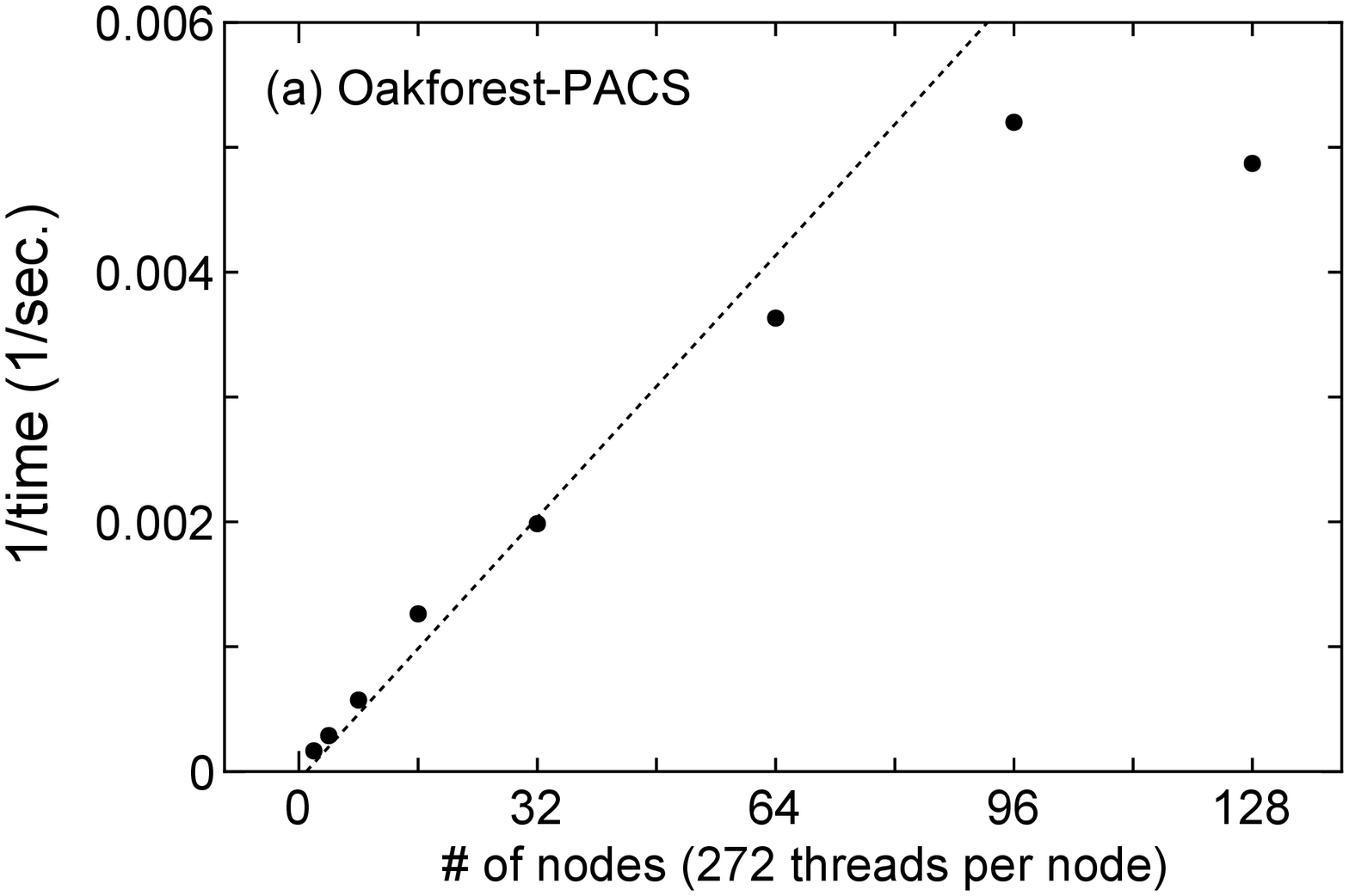}

  \vspace*{3mm}

  \includegraphics[scale=0.4]{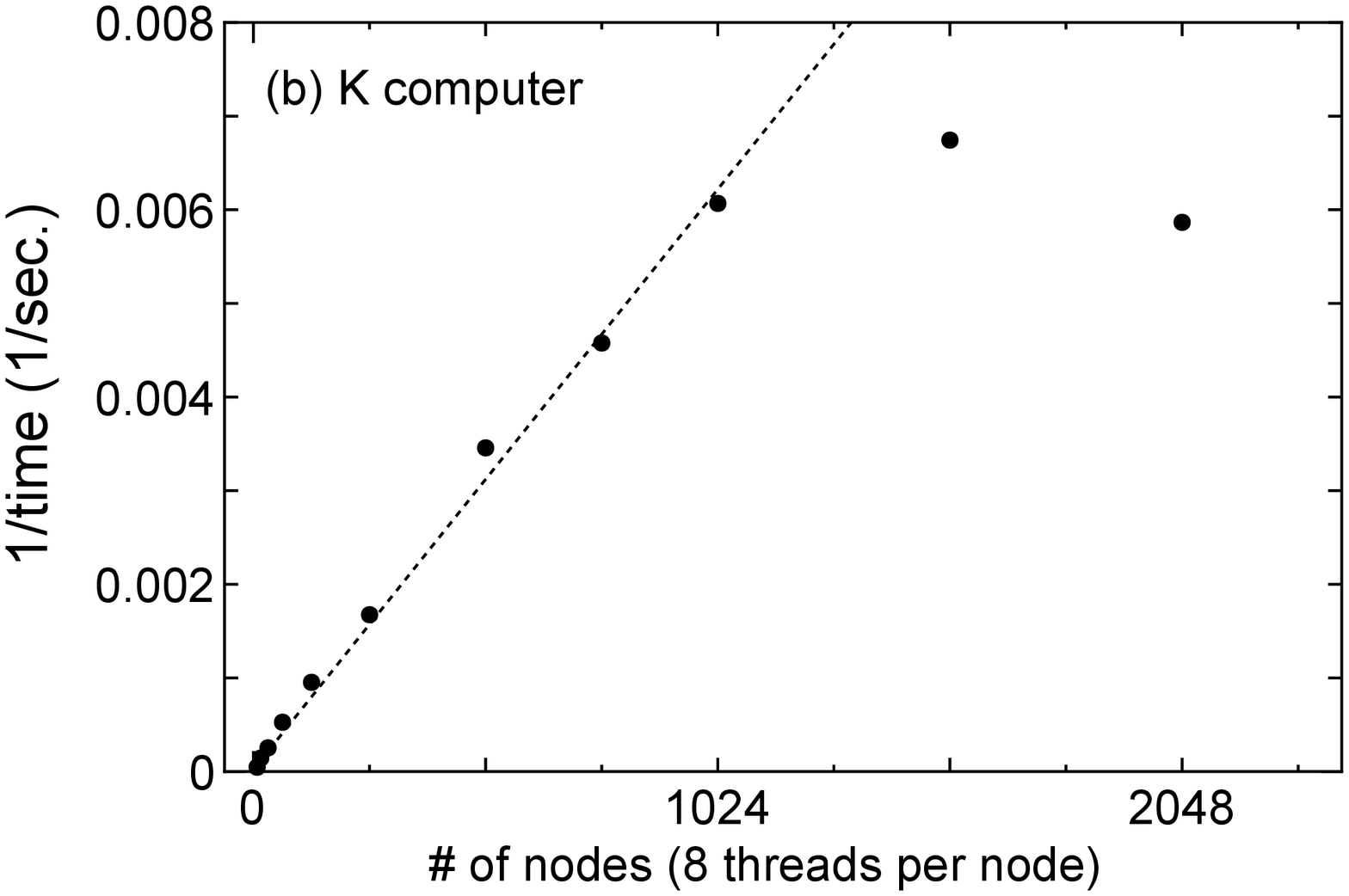}
%  \end{center}
  \caption{
    Parallel performance at (a) Oakforest-PACS \cite{oakforest-pacs}
    and (b) K computer \cite{kcomputer}.
    The elapsed time to obtain the ground-state energy 
    of $^{56}$Ni with the GXPF1A  interaction \cite{gxpf1a}.
    The dotted lines are drawn to guide the eyes.
  }
  \label{fig:parallel-ofp}
\end{figure}

The KSHELL code enables us to perform 
massively parallel computation with hybrid MPI/OpenMP \cite{kshell}.
Figures \ref{fig:parallel-ofp} (a) and (b) show the 
strong scaling of the parallel computation: 
the inverse of the elapsed time to the number of nodes 
for parallel computations.
It shows the elapsed time to obtain the ground-state energy 
of $^{56}$Ni in $pf$ shell,  
the $M$-scheme dimension of which reaches around $1.1\times 10^9$.
These figures  show good parallel performance up to $O(10^4)$
threads and it takes only 192 seconds for the convergence 
utilizing Oakforest-PACS 96 nodes \cite{oakforest-pacs}.

The most time-consuming parts of the LSSM  calculations
are the matrix-vector product and the reorthogonalization
(see Figs.~\ref{fig:time-total-blan} and \ref{fig:time-ne128}).
Especially in parallel computation, 
the elapsed time of the matrix-vector products
exceeds 80\% of the total elapsed time in most cases, 
and it is worth discussing how to compute the matrix-vector
product in parallel 
and its parallel efficiency. 
The parallel computations of the matrix-vector product 
are discussed in Appx.~\ref{sec:perf-matvec} and 
the parallel performance of the reorthogonalization
is discussed in Appx.~\ref{sec:reorth}.

\subsection{Parallel computation of a matrix-vector product}
\label{sec:perf-matvec}

% 28Si 25024860 Non-zero m.e.
% 24Mg 6030189  Non-zero m.e. 28503 dim.,   8099459 line   sparsity 0.7%
% 20Ne 54104 Non-zero m.e. 640 dim., sparsity  13.2%
\begin{figure}[htbp]
  \includegraphics[scale=0.3]{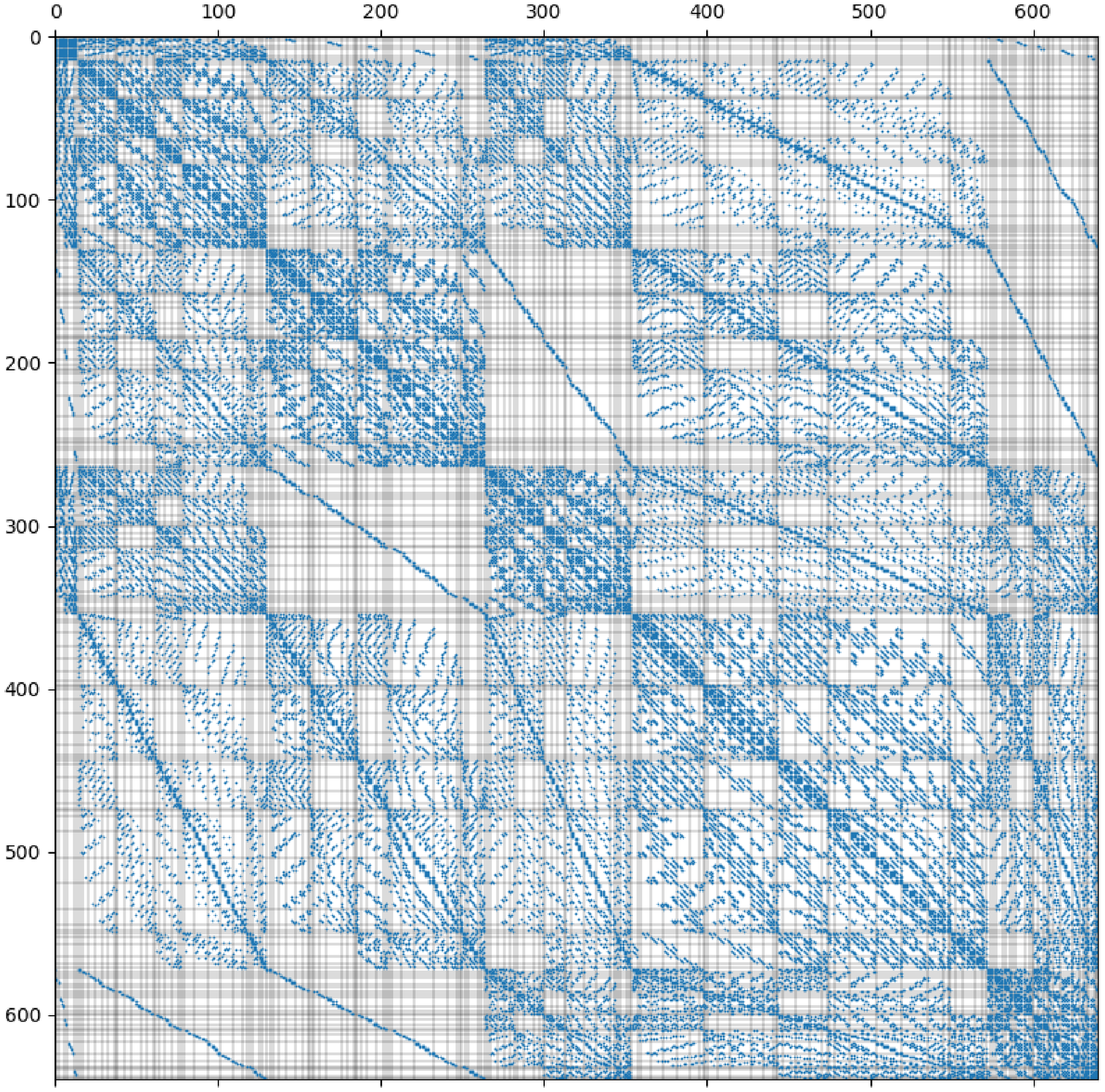}
  \caption{
    Structure of the Hamiltonian matrix for the $M^\pi=0^+$ space
    of $^{20}$Ne with the $sd$-shell model space.
    The only non-zero matrix elements are shown as the blue points.
    The black lines denote the borders of the partitions
    of $(\textbf{N}^{(p)},\textbf{N}^{(n)}, M^{(p)})$.
  }
  \label{fig:Ne20-mat}
\end{figure}

Figure \ref{fig:Ne20-mat} shows the non-zero matrix elements
of the Hamiltonian matrix of $^{20}$Ne with the $sd$-shell model space.
While the $M$-scheme dimension is 640, 
the number of the non-zero matrix elements is 54,104 and thus 
its sparsity is 13.2\%.
The black lines denote the borders of the partition
$(\textbf{N}^\pi,\textbf{N}^\nu, M^\pi)$.
The number of the partitions is 162$\times$162,
which are used as units for the parallel computation,
and we exclude the computation between the partitions that contain 
no matrix elements in advance.
The order of the partitions are shuffled to achieve 
good load balance for practical parallel computations.

\begin{figure}[htbp]
  \includegraphics[scale=0.3]{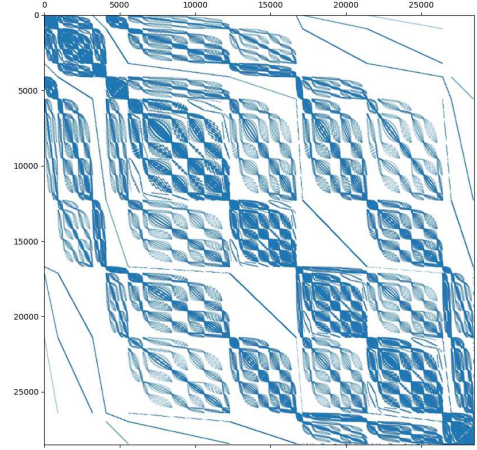}
  \caption{
    Structure of the Hamiltonian matrix for the $M^\pi=0^+$ space 
    of $^{24}$Mg with the $sd$-shell model space.
    The only non-zero matrix elements are shown as the blue points.
  }
  \label{fig:Mg24-mat}
\end{figure}

Figure \ref{fig:Mg24-mat} shows the non-zero matrix elements
of the Hamiltonian matrix of $^{24}$Mg with
the $sd$-shell model space.
While the $M$-scheme dimension is 28,503,  
the number of the non-zero matrix elements is 6,030,189 and thus 
its sparsity is 0.7\%.
The sparsity tends to decrease as the $M$-scheme dimension increases.
The figure represents that the block structure
appears based on the partitions and
block diagonal part is quite dense in comparison with
the non-diagonal part.
This tendency favors more in larger-scale calculations.
Thus, since the product operation of diagonal partitions 
costs far more than that of non-diagonal partitions, 
it is essential to make the operation of diagonal partitions 
equally distributed.
The number of the partitions is 1020, which is not shown
in the figure for simplicity.

The desired features of the algorithm for the parallel computation of the matrix-vector
product are 
\begin{enumerate}
\item Parallel computation based in units of partitions.
\item Good load balance: 
  since diagonal partitions are computationally 
  expensive, they must be distributed equally.
\item The network communications to transfer 
  Lanczos vectors are minimized.
\end{enumerate}

\begin{figure}[htbp]
  \includegraphics[scale=0.3]{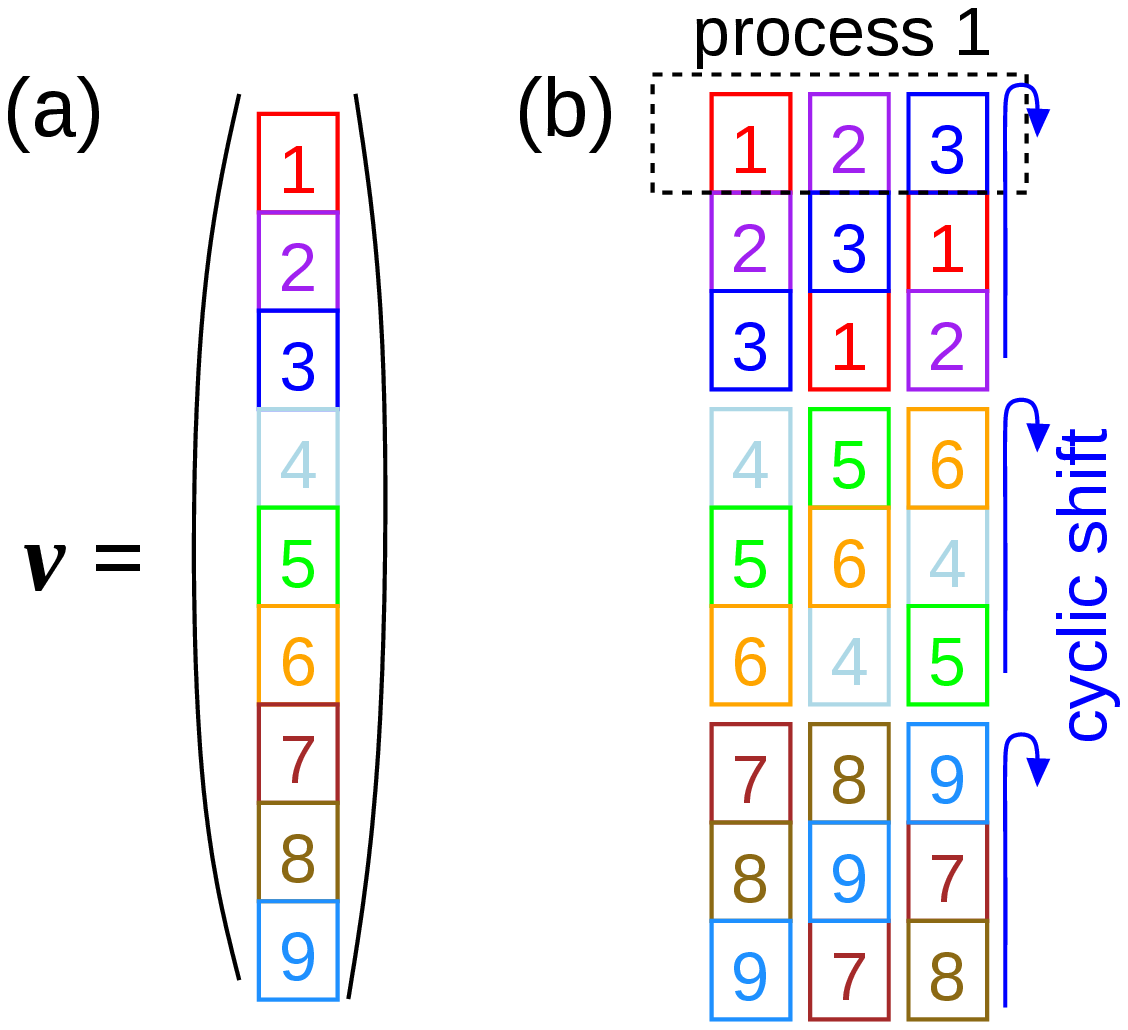}
  \hspace{1.cm}
  \includegraphics[scale=0.3]{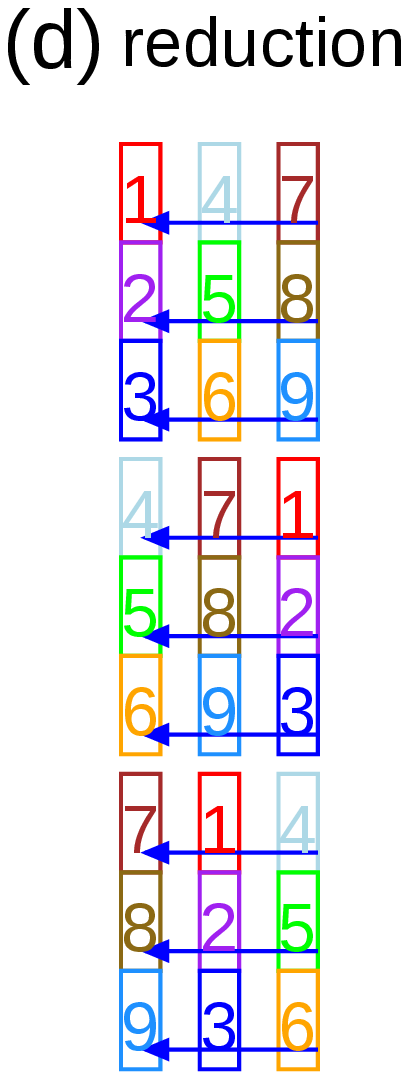}
  \\
  \vspace{0.2cm}
  \includegraphics[scale=0.3]{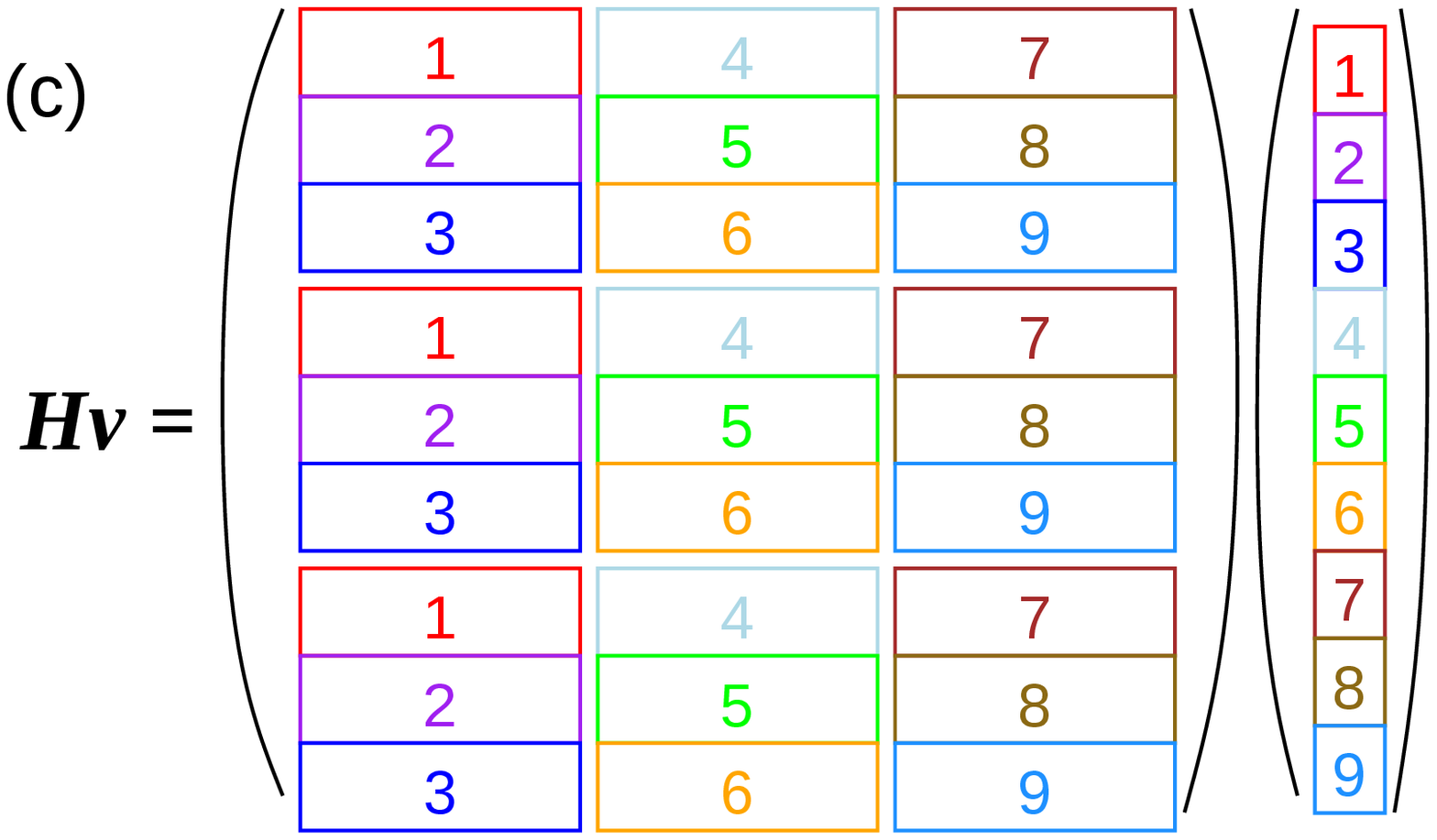}
  \caption{
    Conceptual drawing of the parallel assignment adopted
    in the KSHELL code 
    for a matrix-vector product. 
    (a) A Lanczos vector is distributed over nine processes. 
    (b) Data transfer prepared for the matrix-vector product.
    (c) Assignment on each process for matrix-elements
    generation in a matrix-vector product.
    (d) Data transfer after the generation of matrix elements.
    See text for details. }
  \label{matvec_block}
\end{figure}

In order to satisfy these three features,
% and to reduce the size of network communications,  
we propose a blocked parallel assignment shown
in Fig.~\ref{matvec_block}.
In the figure, we assume nine processes for parallel computation
and the assignment of each process
is shown as colored boxes with
the process number. 

Here, we describe procedures to perform matrix-vector product
in parallel computation.
Firstly, a Lanczos vector is split into nine processes equally
as shown in Fig.~\ref{matvec_block}(a).
In the same manner,
each previous Lanczos vector is split into the processes equally
and is stored on memory.
Secondly, in advance of the matrix-vector product,
the required parts of the vector are
transferred via MPI communications. 
In order to perform the product,
each process needs other parts of the Lanczos vector 
in addition to its own part of the vector.
For example, the first process
requires the first, second, and third parts of the vector, which
are transferred by a cyclic shift as shown in Fig.~\ref{matvec_block}(b).
Thirdly, the matrix-vector product is performed
as shown in Fig.~\ref{matvec_block}(c). 
The Hamiltonian matrix is split into nine parts as shown in
the middle of Fig.~\ref{matvec_block} so that 
dense near-diagonal parts of the matrix are
distributed to each node equally.
Note that 
the boxes in the matrix in the figure represent
the assignments of the on-the-fly generation,
not the store of the elements on memory.
Finally, the reduction of the resultant parts of the vector
is performed via MPI communications shown in Fig.~\ref{matvec_block}(d). 
In total, the size of the network communication at every matrix-vector
product is $16D/\sqrt{N_p}$ bytes where $N_p$ is the number of parallel
nodes and it decreases as $N_p$ increases.
Such two-dimension topology network communication 
matches a torus interconnection network adopted
in K computer.

For comparison, we mention
a simple two-dimensional square-lattice distribution of the matrix. 
It causes
inefficient load balance since
the computation of a part handling the diagonal matrix elements
costs far heavier than others.

\subsection{Parallel computation of  reorthogonalization}
\label{sec:reorth}

The parallel computation of the reorthogonalization is
rather simple: A Lanczos vector is distributed to
all nodes almost equally and the inner product of 
Lanczos vectors is computed in parallel.
Even though the thick-restart algorithm restricts the number
of the Lanczos vectors to be orthogonalized,
the reorthogonalization is the second time-consuming
part of the computation.

\begin{figure}[htbp]
  \includegraphics[scale=0.3]{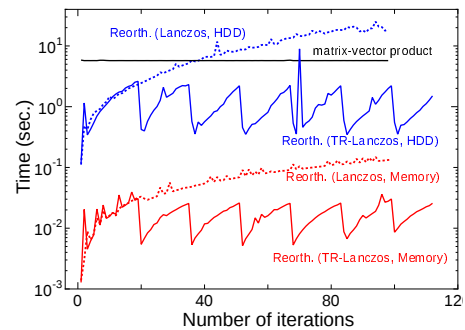}
  \caption{
    Elapsed time of reorthogonalization 
    against the number of the Lanczos iterations
    in comparison with the time of a matrix-vector product
    (black solid horizontal line).
    The red (blue) dotted line shows
    the elapsed time of the reorthogonalization
    of the Lanczos method 
    with storing the Lanczos vectors on memory (on
    HDD). The solid line shows
    those with the thick-restart Lanczos method.
  }
  \label{fig:reorth}
\end{figure}

If the memory capacity is not enough to store the whole
Lanczos vectors (it often occurs in a single-node computation),
these vectors are obliged to be stored in the hard disk drive (HDD)
and to be read at every reorthogonalization, as done in
conventional shell-model codes such as MSHELL64.
As the number of iterations increases, 
the cost of the disk I/O of the HDD grows and
it overcomes that of the 
matrix-vector product as shown in Fig.~\ref{fig:reorth}.
The block algorithm tends to make the number of the Lanczos
vectors larger and gets worse.

Figure \ref{fig:reorth} shows the
time of the reorthogonalization as a function of the
number of the Lanczos iterations in the simple Lanczos method
utilizing the HDD.
The elapsed time was measured
to obtain the ground-state energy of
the $^{56}$Ni in the $pf$-shell model space
at K computer with 480 nodes \cite{kcomputer}.
The time of the reorthogonalization using the HDD 
surpasses the
time of the matrix-vector product,
the black solid line in Fig.~\ref{fig:reorth}, 
at the 39th iteration and becomes the bottleneck.
By using the thick-restart method in which the restart process 
is done when the number of the Lanczos vectors reaches 40,
the time of the reorthogonalization (the solid blue line)
becomes much smaller than that of the matrix-vector product.
However, the thick restart makes the convergence slightly slow
as discussed in Sect.~\ref{sec:conv}.
Some irregular spikes in the figure would be caused 
by other jobs running at the K computer.

When we store whole the Lanczos vectors on memory,
the time of the reorthogonalization is much reduced 
and is shown as the red lines in Fig.~\ref{fig:reorth}.
In this case, the time of the reorthogonalization
is two orders of magnitude smaller than that of a
matrix-vector product, but the memory capacity
is required to keep the whole vectors.

%-----------------------------------------------------------------------

\end{document}